\documentclass[10pt]{article}
\usepackage{amssymb,latexsym,amsmath,amsthm,graphicx,subfigure}
\usepackage{amscd}
\usepackage{epsf}

\makeatletter \@addtoreset{figure}{section}
\def\thefigure{\thesection.\@arabic\c@figure}
\def\fps@figure{h, t}
\@addtoreset{table}{bsection}
\def\thetable{\thesection.\@arabic\c@table}
\def\fps@table{h, t}
\@addtoreset{equation}{section}

\newif\ifamsfonts\amsfontstrue \ifamsfonts
\font\twlbbb=msbm10 scaled\magstep1 \font\egtbbb=msbm8
\font\sixbbb=msbm6
\newfam\mathbbfam
\textfont\mathbbfam=\twlbbb \scriptfont\mathbbfam=\egtbbb
\scriptscriptfont\mathbbfam=\sixbbb
\makeatother

\newtheorem{thm}{Theorem}[section]\newtheorem{prop}[thm]{Proposition}
\newtheorem{lem}[thm]{Lemma}

\newtheorem{remark}{Remark}[section]
\newtheorem{example}{Example}[section]
\newcommand{\R}{\mathbb{ R}}
\newcommand{\h}{\mathfrak{ h}}
\newcommand{\g}{\mathfrak{ g}}
\newcommand{\D}{\mathfrak{ d}}
\newcommand{\N}{\mathcal{ N}}
\newcommand{\M}{\mathcal{ M}}

\newcommand{\m}{\mathfrak m}
\newcommand{\w}{\mathfrak w}
\newcommand{\I}{\mathcal I}
\newcommand{\A}{\mathcal A}
\DeclareMathOperator{\pr}{pr}
\DeclareMathOperator{\Span}{span}
\DeclareMathOperator{\diag}{diag}
\DeclareMathOperator{\ad}{ad}

\begin{document}

\title{Integrable nonholonomic geodesic
flows on compact Lie groups \footnote{AMS Subject Classification
37J60, 37J35, 70H45}}
\author{Yuri N. Fedorov \\
 Department of Mathematics and Mechanics
 \\ Moscow Lomonosov University, Moscow, 119 899, Russia \\
{\footnotesize e-mail: fedorov@mech.math.msu.su} \\
and \\
 Departament de Matem\`atica I, \\
Universitat Politecnica de Catalunya, \\
Barcelona, E-08028 Spain \\
{\footnotesize e-mail: Yuri.Fedorov@upc.es} \\
and \\
Bo\v zidar Jovanovi\' c \\
Mathematical Institute, SANU \\
Kneza Mihaila 35, 11000, Belgrade, Serbia  \\
{\footnotesize e-mail: bozaj@mi.sanu.ac.yu } }
\maketitle

\begin{abstract} This paper is a review of recent results on integrable
nonholonomic geodesic flows of left--invariant metrics and left-
and right--invariant constraint distributions on compact Lie groups.
\end{abstract}

\tableofcontents

\section{Introduction}

This paper is a review of recent results on
integrable flows on compact Lie groups under nonholonomic constraints.
We mostly follow papers \cite{FeKo, Jo1, Jo2, FeJo},
trying to present their results within a unified framework.
Furthermore, some new examples of integrable nonholonomic systems are given.

\subsection{Nonholonomic Geodesic Flows}
We start with basic definitions and settings.
Let $(Q,ds^2)$ be $n$--dimensional
Riemannian manifold $Q$ with a nondegenerate matric $ds^2$and
a Levi--Civita connection $\nabla$, $D$ be
a nonintegrable $k$--dimensional distribution on the tangent
bundle $TQ$. A smooth path $\gamma(t)\in Q,\; t\in\Delta$ is
called {\it admissible} (or allowed by constraints) if  the
velocity $\dot\gamma(t)$ belongs to ${D}_{\gamma(t)}$ for all
$t\in\Delta$. There are two approaches to define geodesic lines
among admissible paths: by induced connection as ``straightest''
lines  and by the variation principle as ``shortest'' lines. We
shall deal with the first approach which arises from mechanics.

The admissible path $\gamma(t)$ is called {\it a nonholonomic
geodesic\/} if it satisfies the equations
\begin{equation}
\pi(\nabla_{\dot\gamma(t)}\dot \gamma(t))=0,
\label{geodesics}
\end{equation}
where $\pi: T_q Q\to {D}_q$, $q\in Q$ is the orthogonal projection.

Equivalently, we can introduce the Lagrangian function $l=\frac12
(K\dot q,\dot q)$, where $K$ is the metric on $Q$ also regarded as
a mapping $K: TQ\to T^*Q$.
Let $q=(q_1,\dots,q_n)$ be some local coordinates on $Q$.
The trajectory of the system $q(t)$ that
satisfies the constraints is a solution to the
Lagrange--d'Alambert equations \begin{equation}
\left( \frac{\partial l}{\partial q} - \frac{d}{dt}\frac{\partial
l}{\partial \dot q},\eta\right)=
\sum_i \left(\frac{\partial l}{\partial q_i} -
\frac{d}{dt}\frac{\partial l}{\partial \dot q_i}\right)\eta_i=0,
\quad \mathrm{for\;all} \quad \eta\in D_q. \label{Lagrange}
\end{equation}

One can also write the Lagrange-d'Alambert equations as a
first-order system on the $(n+k)$-dimensional {\it constraint
submanifold} $\M=K(D)$ of the cotangent bundle $T^*Q$.
Let $D$ be locally defined by
$\rho=n-k$ independent 1-forms $\alpha^i$
$$
D_q=\{ \xi\in T_q Q, \; (\alpha^j_q,\xi)=\sum_i\alpha_i^j\xi_i=0,\; j=1,\dots,\rho \}.
$$
Then $\M$ is locally given by the equations $(\alpha^i_q,K^{-1}_q p)=0$,
$i=1,\dots,\rho$.
Let $p_i={\partial l}/\dot q_i$, $i=1,\dots,n$ be momenta which together with $q$
provide canonical coordinates on $T^*Q$.
Let $h(q,p)=\frac12 (p,K^{-1}_q p)$ be the
Hamiltonian function (the usual Legendre transformation of $L$).
The equations (\ref{Lagrange}) are equivalent to
\begin{equation}
\dot p_i=-\frac{\partial h(q,p)}{\partial q_i}+  \sum_{i=1}^{\rho}
\lambda_j \alpha^j(q)_i, \quad \dot q_i=\frac{\partial
h(q,p)}{\partial p_i}, \qquad
i=1,\dots,n,\label{Hamilton}
\end{equation}
where Lagrange multipliers are chosen such that the solutions
$(q(t),p(t))$ belong to $\M$.

As for the Hamiltonian systems, the Hamiltonian function is always
the first integral of the system. There is also a nonholonomic
version of the Noether theorem (see \cite{KK, AKN, FeKo, BKMM}).

\paragraph{The Noether theorem.}
Suppose that a Lie group $\mathfrak G$ acts on the configuration space
$Q$ and that the action is naturally extended to $TQ$ and $T^*Q$.
The momentum mappings $\Psi_l: TQ\to \g^*$ and $\Psi^*:
T^*Q\to\g^*$ are defined by
\begin{equation}
\Psi_l(q,\dot q\, \vert \xi)=\left(
\frac{\partial l}{\partial \dot q},\xi_Q \right)
=(K_q \dot q,\xi_Q), \quad \Psi^*(q,p\, \vert \xi)=(p,\xi_Q),
\label{moment_map}
\end{equation}
where $\xi_Q$ is the vector field on $Q$ associated to the action
of one-parameter subgroup $\exp(t\xi)$, $\xi\in\g=T_{Id}\mathfrak G$.

\begin{thm}
Assume that $\xi_Q$ is a section of the distribution $D$ and the
one-parameter subgroup $\exp(t\xi)$ preserves $l$ (or $h$). Then
$\Psi_l(\xi)$ is the first integral of the system
(\ref{Lagrange}), or equivalently, $\Psi^*(\xi)$ is the first
integral of (\ref{Hamilton}).
\end{thm}

\paragraph{Invariant measure and integrability.}
The equations (\ref{Hamilton}) are not Hamiltonian.
This is why it is  still not clear how to define the notion of
complete integrability for nonholonomic systems (see \cite{BaCu}).
However, in some cases they have an invariant measure,
a rather strong property, which puts the system close to Hamiltonian systems.
In particular, if apart from the Hamiltonian there exist $\dim\M-3$ additional
independent integrals, then, by the Euler--Jacobi theorem, the solutions of
(\ref{Hamilton}) can be found by quadratures.

The importance of an invariant measure
for integrability of nonholonomic systems was
indicated by Kozlov in \cite{Koz1}, where various examples
were discussed (see also \cite{AKN}).
Namely, consider a non-Hamiltonian system
\begin{equation}
\dot x=f(x), \qquad x\in\mathbb{R}^m,
\label{d1}
\end{equation}
having an invariant measure $\mu(x)\,dx$
and $m-2$ first integrals $F_1(x),\dots,F_m(x)$.
If the latter are independent on the invariant set
$M_c=\{x\in\mathbb{R}^m,\; F_i(x)=c_i, \; i=1,\dots,m-2\}$, then
$M_c$ is a two-dimensional submanifold and the flow on
$M_c$ has also an invariant measure. Then, accoring to the Euler--Jacobi theorem,
solutions of  (\ref{d1}) lying on $M_c$ can be found by quadratures.
Moreover, if $L_c$ is a compact connected component of $M_c$
and $f(x) \ne 0$ on $L_c$, then, if orientable, $L_c$ is diffeomorphic
to two-dimensional torus.

According to Kolmogorov's theorem
on reduction of differential equations with a smooth invariant measure on
a torus (\cite{Kol}),
one can find angular coordinates $\varphi_1,\varphi_2$ on $L_c$, in which the reduction
of equations  (\ref{d1}) takes the form similar as in the Liouville theorem:
$$
\dot\varphi_1=\frac{\Omega_1}{\Phi(\varphi_1,\varphi_2)}, \quad
\dot\varphi_2=\frac{\Omega_2}{\Phi(\varphi_1,\varphi_2)},
$$
where $\Omega_1,\Omega_2$ depend on the constants of motion $c_1,\dots,c_{m-2}$ only
and $\Phi$ is a smooth positive $2\pi$--periodic function in
$\varphi_1,\varphi_2$, the density of the induced invariant measure on $L_c$.

Therefore, it is natural to call the system (\ref{d1})
{\it completely integrable} if it can be integrated by the Jacobi theorem; or,
more generally (see \cite{VeVe1, VeVe2}), if the phase space is almost
everywhere
foliated by invariant tori $\mathbb{T}^k\{\varphi_1,\dots,\varphi_k\}$ with the
dynamics of the form
\begin{equation}
\dot \varphi_1=\frac{\Omega_1}{\Phi(\varphi_1,\dots,\varphi_k)},\quad \dots,
\quad \dot \varphi_k=\frac{\Omega_k}{\Phi(\varphi_1,\dots,\varphi_k)}.
\label{quasi-periodic}
\end{equation}

The above definition of complete integrability is slightly
different from the definition of complete integrability of
non-Hamiltonian systems given in \cite{Bo2, Zu}. Namely, here we
have quasi-periodic motions after the time substitution $d\tau=\Phi^{-1}(\varphi)dt$.

The existence of an invariant measure for smooth dynamical systems
and for a class of nonholonomic systems with symmetries is studied
in \cite{Kozlov} and \cite{Bl_Z3}, respectively. Various
mechanical examples with an invariant measure can be found in
\cite{BM2}. The authors of the paper \cite{VeVe1, VeVe2} constructed
nonholonomic systems on unimodular Lie groups with right-invariant
nonintegrable constraints and a left-invariant metric (so called
{\it LR systems}), and showed that they always possess an
invariant measure, whose density can be effectively calculated. In
particular, the motion of a rigid body  around a fixed point under
a nonholonomic constraint (projection of the angular velocity to
the fixed vector in space is constant) is described  by an
integrable LR system (\cite{VeVe1}). Similar integrable problems on
Lie groups with left--invariant constraints are studied in
\cite{FeJo, Jo1, Jo2}.  Also, an important example of an
integrable nonholonomic mechanical system, the problem of rolling of a
homogeneous ball on a surface of revolution (the Routh problem),
was treated in detail in \cite{He, Ze1}.

\subsection{Chaplygin Systems}
Another approach to the integrability of nonholonomic systems
is based on their reduction to a Hamiltonian form
after an appropriate time rescaling. First, following \cite{Koi} and \cite{BKMM},
let us recall some basic facts about the Chaplygin systems.

Let $(Q,l,D)$ be a nonholonomic system with a Lagrangian $l$
of the natural mechanical type, with kinetic energy that correspods to
the metric $ds^2$ and the potential function $v$.
Assume that there is a bundle structure $\pi: Q\to N$ with the base
manifold $N$ and let the map $\pi$ be a submersion,
such that $T_q Q= D_q\oplus V_q$ for all $q$.
Here $V_q$ is the kernel of $T_q\pi$ called the {\it vertical space} at $q$.
Then the distribution $D$ can be seen as a collection of {\it horizontal spaces}
of the  {\it Ehresmann connection} associated to $\pi: Q\to N$.
Given a vector $X_q\in T_q Q$, there is a decomposition $X_q=X_q^h+X_q^v$, where
$X_q^h\in D_q$, $X_q^v\in V_q$.
The {\it curvature} of the connection is the vertical valued 2-form $B$ on $Q$
defined by
$$
B(X_q,Y_q)=-[\bar X_q^h,\bar Y_q^h]_q^v ,
$$
where $\bar X$ and $\bar Y$ are smooth vector fields on $Q$ obtained by
extending of $X_q$ and $Y_q$.

By applying the Ehresmann connection the Lagrange--d'Alambert equations
(\ref{Lagrange}) can be represented in the form (see \cite{BKMM})
\begin{equation}
\left(\frac{\partial l_c}{\partial q}-\frac{d}{dt}\frac{\partial l_c}{\partial
\dot q},\eta\right)=
\left(\frac{\partial l}{\partial \dot q},B(\dot q,\eta)\right),\quad
\mathrm{for\;all} \quad \eta\in D_q,
\label{Chaplygin}
\end{equation}
where $l_c(q,\dot q)=l(q,\dot q^h)$ is the {\it constrained Lagrangian}.

Now, suppose that $\pi: Q\to N=Q/{\mathfrak G}$ is a principal bundle
with respect to the
{\it left\/} action of a Lie group ${\mathfrak G}$, and $D$ is a principal
connection, i.e., $D$ is a $\mathfrak G$-invariant distribution.
Let  the Lagrangian $l$ be also $\mathfrak G$-invariant, i.e.,
$\mathfrak G$ acts by isometries on Riemannian manifold $(Q,ds^2)$ and
$v$ is a $\mathfrak G$--invariant function.
Then the constrained Lagrangian $l_c$
induces a well defined reduced Lagrangian $L: TQ\to\R$ via identification
$TN\approx D/{\mathfrak G}$.
The reduced Lagrangian $L$ is of the natural mechanical type as well.
Its kinetic energy is given by metric $ds^2_D$ and its potential energy will be
denoted by $V$.

Under the above assumtions, equations (\ref{Chaplygin}) are $\mathfrak G$-invariant and
induce  {\it reduced Lagrange--d'Alambert} equations on the tangent bundle $TN$,
\begin{equation}
\left(\frac{\partial L}{\partial q}-\frac{d}{dt}\frac{\partial L}{\partial
\dot q},\eta\right)=\Sigma(\dot q,\eta),\quad
\mathrm{for\;all} \quad \eta\in T_q N.
\label{Chaplygin_red}
\end{equation}
Here $\Sigma$ is semi-basic two-form given by the right hand side
of (\ref{Chaplygin}) and $q=(q_1,\dots,q_k)$ are some local
coordinates on the base space $N$. From (\ref{moment_map}) we see
that $\Sigma$ depends on the curvature of the connection $D$ and
on the momentum mapping $\Phi_l$.

The system $(Q,l,D,\mathfrak G)$ is referred to as {\it a (generalized) Chaplygin system}
(see \cite {Koi, BKMM}), as a generalization of classical Chaplygin systems
with Abelian symmetries \cite{Ch}.

\begin{remark}
\label{submersions}{\rm Note that horizontal and vertical spaces
do not need to be orthogonal with respect to the metric $ds^2$. In
fact,  if $D$ is $ds^2$-orthogonal to the leaf of $\mathfrak G$-action,
then $D$ will be an invariant submanifold of the
nonconstrained geodesic flow of the metric $ds^2$, and the right
hand sides of  equations (\ref{Chaplygin}) will be zero.
In this case, $ds^2_D$ coincides with the {\it submersion metric} induced from $ds^2$.}
\end{remark}

\paragraph{Chaplygin's reducing multiplier.}
Let $p_i=\partial L/\partial \dot q_i$, $i=1,\dots,k$ be momenta,
$g_{ij}$ the metric tensor of $ds^2_{D}$ and $g^{ij}$ the dual
metric on $T^*N$. Then the reduced Lagrangian has the form $L(q,\dot q)=\frac12
\sum g_{ij} \dot q_i\dot q_j -V(q)$. We also introduce the
Hamiltonian function $H(q,p)=\frac12 \sum g^{ij} p_ip_j+V(q)$.
The reduced system (\ref{Chaplygin_red}) can be rewritten as
a first-order dynamical system on $T^*N$:
\begin{equation}
\dot q_i=\frac{\partial H}{\partial p_i},\quad
\dot p_i=-\frac{\partial H}{\partial q_i}+\Pi_i(q,p), \qquad i=1,\dots,k.  \label{2.1}
\end{equation}
The functions $\Pi_i$ are quadratic in momenta and can be regarded
as non-Hamiltonian perturbations of the equations of motion of a
particle on $N$.

Let $\Omega=\sum dp_i\wedge dq_i$ be the standard symplectic form
on $T^*N$. The equations (\ref{2.1}) have an invariant measure
$f\Omega^k$ if
$\sum_i \left( \frac{\partial (f\dot q_i)}{\partial q_i} +
\frac{\partial(f\dot p_i+f\Pi_i)}{\partial p_i} \right) =0$.
Since the finction $f$ depends only on the coordinates $q$, this is equivalent to
condition
\begin{equation}
d(\ln f)+\alpha=0, \label{x3}
\end{equation}
where the one-form $\alpha$ is given by
$\sum_i\frac{\partial\Pi_i}{\partial p_i}\vert_{\dot
q=gp}=(\alpha,\dot q)$.

\begin{remark} {\rm The paper \cite{St} (see also \cite{CCLM}) contains
a nontrivial observation about the density of the invariant
measure, which in our terms reads as follows. Suppose that system
(\ref{2.1}) has an invariant measure with density $f(q,p)$ in the case
of absence of potential ($V(q)=0$). Then one can check
that the function $f_0(q)=f(q,0)$ is also a solution of
(\ref{x3}). In other words, if the reduced system (\ref{2.1}) has
an invariant measure for $V=0$, one can take this measure to be of
the form $f(q) \Omega^k$. Then, since (\ref{x3}) does not depend
on the potential, the reduced system (\ref{2.1}) has the same
invariant measure in the presence of a potential field $V(q)$ as well.}
\end{remark}

Now consider time substitution $d\tau={\mathcal N} (q)dt$, where
${\mathcal N}(q)$ is a differentiable nonvanishing function on
$Q$, and denote $q^{\prime}={dq}/{d\tau}$. Then we have the
following commutative diagram
\begin{equation*}
\begin{CD}
 TN\{q,\dot q\} @ > q^{\prime}=\dot q/{\mathcal N}(q) >> TN\{q,q^{\prime}\} \\
 @ V p=g \dot q VV   @ VV \tilde p={\mathcal N}^2 g q^{\prime} V \\
 T^*N\{q,p\}  @ >  \tilde p=\mathcal{\ N} p >>  T^*N\{q,\tilde p\} .
\end{CD}
\end{equation*}

The Lagrangian and Hamiltonian functions in the coordinates
$\{q,q^{\prime}\} $ and $\{q,\tilde p\}$ take the form
\begin{equation*}
L^*(q,q^{\prime})= \frac{1}{2}\sum \mathcal{\ N}^2 g_{ij}
q_i^{\prime}q_j^{\prime}-V(q), \quad H^*(q,\tilde p)=\frac12\sum \frac{1}
{\mathcal{\ N}^2} g^{ij} \tilde p_i \tilde p_j+V(q).
\end{equation*}

There is a remarkable relation between the existence of an invariant measure
of the reduced system (\ref{2.1}) and its reducibility to a Hamiltonian form
(see \cite{FeJo}).

\begin{thm} \label{Th_reduce} \begin{description}
\item{1).}
Suppose that after the time substitution $d\tau=\N(q) dt$
the equations (\ref{2.1}) become Hamiltonian,
\begin{equation}
q_i'=\frac{\partial H^*}{\partial \tilde p_i}, \quad
\tilde p_i'= -\frac{\partial H^*}{\partial q_i}\, .
\label{x2}
\end{equation}
Then the function $f(q)=\N(q)^{k-1}$ satisfies the equation
(\ref{x3}), i.e., the original system (\ref{2.1}) has the
invariant measure with density $f(q)$.

\item{2).} For $k=2$, the above statement can also be inverted:
the existence of the invariant measure with the density $\N(q)$
implies that in the new time $d\tau=\N(q)dt$, the system
(\ref{2.1})  gets the Hamiltonian form (\ref{x2}).
\end{description}
\end{thm}

In nonholonomic mechanics the factor $\N$ is known as the {\it
reducing multiplier\/}, item 2) of this theorem is referred to  as
{\it Chaplygin's reducibility theorem\/} (see \cite{Ch2, Ch} or
section III.12 in \cite{NeFu}).
 Notice that for $k>2$, the multiplier $\N (q)$ and the density of
the invariant measure of system (\ref{2.1})  do not coincide.
Also, the existence of the multiplier do not depends on the
potential $V$.

There are many examples  of the Chaplygin reducing multiplier for $k=2$.
Since many conditions on the metric and constraints are imposed,
until recently there were no nontrivial examples of multidimensional systems,
appart of several examples for $k=3,4$ with the property that factor
$\N(q)$ depends only on one coordinate,
that are reducible to a Hamiltonian form by the Chaplygin procedure
(\cite{NeFu, Efimov, Iliev, Moschuk}).

As an alternative, in the reduction of Chaplygin systems
one can use the symplectic (or Poisson) framework (see \cite{St,BS,
CCLM, Co}). Such systems can be represented in a Hamilton-like
form with respect to an nondegenerate (almost-symplectic) 2-form,
which however may be not closed Namely, let $\Xi$ be the Legandre
transformation of the semi-basic form $\Sigma$. Then one can
write (\ref{2.1}) as
$$
\Omega_{nh}(X_H,\cdot)=dH(\cdot),\quad \rm{where}\quad  \Omega_{nh}=\Omega+\Xi.
$$
In this framework, the Chaplygin multiplier is a function $\N$
such that the form $\tilde\Omega=\N\Omega_{nh}$ is closed. Then,
after rescaling $Y=X/\N$, we obtain the Hamiltonian system
$\tilde\Omega(Y,\cdot)=dH(\cdot)$ (see \cite{St, He, CCLM, EKR}).
Contrary to the procedure described in Theorem \ref{Th_reduce},
here the vector field $Y$ has no direct mechanical description.

Recently, necessary and sufficient conditions for the existence of
an invariant measure of the reduced system in case when the
Lagrangian of the system is of a pure kinetic energy type are
given in \cite{CCLM, Co}.

\subsection{Contents of the Paper}
In section 2 we consider the systems with left--invariant metrics
and left--invariant constraint distributions, so called {\it LL systems}.
The equations of the motion reduce to the Euler--Poincar\'e--Suslov
equations on the corresponding Lie algebra.
Although such equations generally are not Hamiltonian, their nice algebraic structure
allows us to construct various integrable examples with an invariant measure.

In section 3 we consider a class of LR systems (left--invariant metrics
and right--invariant constraint distributions),
which can be regarded as Chaplygin systems on the
principle bundle $G \to Q=G/H$, $H$ being a Lie subgroup.
We show that, in contrast to generic Chaplygin systems, the reductions of our LR
systems onto the homogeneous space $Q$ always possess an invariant measure.
Then we study the case $G=SO(n)$, when LR systems are
multidimensional generalizations of the Veselova problem of a nonholonomic rigid
body motion, which admit a reduction to the system with an invariant measure on
the (co)tangent bundle on the unit sphere $S^{n-1}$.
For a special choice of the left-invariant metric on $SO(n)$, we prove that
under a time reparameterization, the reduced system becomes an integrable Hamiltonian system
describing a geodesic flow on the unit sphere $S^{n-1}$. This provides a first
multidimensional example of a nonholonomic system for which the
celebrated Chaplygin reducibility theorem is applicable.
Lastly, we present an explicit reconstruction of the motion on the group
$SO(n)$.

Finally, in section 4 we present another class of systems on an unimodular Lie
group $G$,
which always possess a non-trivial invariant measure and which are obtained as
modifications of a geodesic flow on $G$ with respect to a sum of a left- and a
right-invariant metrics, so called L+R systems. It appears that
a nonholonomic LR system on a group $G$ can be obtained
as a limit case of an appropriate L+R system on this group. As an example, we
consider a nonholonomic mechanical system called the spherical support.

\section{LL Systems}
\subsection{Euler--Poincar\'e--Suslov Equations}
In this section we consider nonholonomic systems $(G,l,D)$
with a left-invariant distributions $D$ and a left-invariant Lagrangians $l$
that describes left-invariant metrics on a compact connected Lie group $G$.
Let $\g=T_{Id}G$ be the Lie algebra of $G$.
In what follows we shall identify
$\g$ and $\g^*$ by $Ad_G$ invariant scalar product
$\langle \cdot,\cdot\rangle$, and $TG$ and $T^*G$ by bi-invariant metric on $G$.
For clearness, we shall use the symbol $\omega$ for the elements in $\g$
and the symbol $x$ for the elements in $\g^*\cong \g$.

Let
$$
\D=\{\omega\in G,\; \langle \omega,a^i \rangle =0, \; i=1,\dots,\rho \}\subset \g
$$
be the restriction of the left-invariant distribution $D$  to the
algebra, for some constant and linearly independent vectors $a^i$
in $\g$. From the left invariance condition we have $D_g=g\cdot
\D$. The distribution is nonintegrable if and only if $\D$ is not
a subalgebra. Also, it is sufficient to give a Lagrangian at one
point of the group, for instance the identity $l(g,\dot
g)=\frac12\langle \I\omega,\omega \rangle$, $\omega=g^{-1}\cdot
\dot g$. Here $\I: \g\to \g$ is a symmetric positive definite
(with respect to $\langle \cdot,\cdot\rangle$) operator. The
Hamiltonian in the left-trivialization is given by $H(x)=\frac12
\langle \A(x),x\rangle$, $\A=\I^{-1}$. The corresponding
left-invariant metric will be denoted by $ds^2_\I$.

Let $\m$ be the restriction of the constraint  submanifold $\mathcal M$ to $\g$,
that is $\m=\I(\D)$.
Equations (\ref{Hamilton}) are $G$--invariant and reduce to $\m$,
\begin{equation}
\dot x = [x,\nabla H(x)]+ \sum_{i=1}^{\rho} \lambda_i a^i=[x, \A(x)] +
\sum_{i=1}^{\rho} \lambda_i a^i,
\label{EPS}
\end{equation}
where $\lambda_i$ are Lagrange multipliers chosen such that $x$ belongs to
$\m=\I(\D)$, i.e.,  such that $\omega=\A(x)$ belongs to $\D$:
$\langle \A(x),a^i \rangle =0$, $i=1,\dots, \rho$.
In other words, the following commutative diagram holds
\begin{equation}
\begin{CD}
{\mathcal M} @ > {\mathcal P}^t  >> {\mathcal M} \\
 @ V \Lambda  VV @  VV  \Lambda V \\
 \m @ > P^t >> \m  ,
\end{CD}
\label{diagram}
\end{equation}
where ${\mathcal P}^t$  and $P^t$ are phase flows of the nonholonomic geodesic
flow and the system (\ref{EPS}) respectively,
and $\Lambda$ maps $g\cdot x \in T_g G$ to $x\in \g$.

Following \cite{FeKo}, we shall call  (\ref{EPS})   the {\it
Euler--Poincar\' e--Suslov (EPS) equations}, as a generalization
of the Suslov problem of the nonholonomic rigid body motion (see
the example below).

These equations have a quite different nature in comparison
with the Euler--Poincar\' e equations $\dot x=[x,\A(x)]$.
In particular, as indicated in \cite{Koz2},
in the case of only one constraint $\langle a,\A(x)\rangle =0$,
they have a smooth invariant measure if and only if $[a,\A(a)]=\mu a$.

\paragraph{Reconstruction of the motion on the group.}
In the Hamiltonian case, the integrability of the reduced system implies
generally a non-commutative integrability of the original system, namely
the phase space is foliated by invariant isotropic tori with quasi-periodic
dynamic (see \cite{Zu}). However there is no such analog in the nonholonomic setting.
To reconstruct the motion $(g(t),\dot g(t))$ on the whole phase space,
we have to solve the kinematic equation
$$
g^{-1}(t)\cdot \dot g(t)=\omega(t)=\A(x(t)),
$$
where $x(t)$ are solutions of (\ref{EPS}), i.e., to find all trajectories
in $\mathcal M$ that projects to the given trajectory $x(t)$ in $\mathfrak m$.
In particular, if $x(t)$ is a relative equilibrium ($x(t)=x(t_0)$ for all $t$)
or if $x(t)$ is a relative periodic orbit ($x(t+T)=x(t)$ for all $t$), then
the invariant set $\Lambda^{-1}(\{x(t), t\in\R\})\subset\mathcal M$
is foliated by invariant tori of maximal dimension $\mathrm{rank}\,G$
or $\mathrm{rank}\,G+1$, respectively (e.g., see \cite{He}).

\paragraph{Multidimensional Suslov problem.}
The most natural example of LL systems is the  nonholonomic Suslov problem,
which describes the motion of an $n$-dimensional rigid body with a fixed point,
that is, the motion on the Lie group $SO(n)$, with certain left-invariant
nonholonomic constraints.

For a path $g(t)\in SO(n)$, the angular velocity of the body is
defined as the left-trivialization $\omega(t)=g^{-1}\cdot g(t) \in
so(n)$.  The matrix $g\in SO(n)$ maps a coordinate system fixed in
the body to a coordinate system fixed in the space. Therefore, if
$e_1=(e_{11},\dots,e_{1n})^T\dots,e_n=(e_{n1},\dots,e_{nn})^T$ is
the  orthogonal frame of unit vectors fixed in the space {\it and
regarded in the moving frame}, we have
$$
E_1=g\cdot e_1,\; \dots,\;  E_n=g\cdot e_n,
$$
where $E_1=(1,0,\dots,0)^T,\; \dots,\; E_n=(0,\dots,0,1)^T$. From
the conditions $0=\dot E_i=\dot g \cdot e_i+g\cdot \dot e_i$, we
find that the vectors $e_1,\dots,e_n$ satisfy the Poisson
equations
\begin{equation}
\label{Poisson}
\dot e_i= - \omega e_i, \qquad  i=1,\dots,n.
\end{equation}

The left-invariant metric on $SO(n)$ is given by non-degenerate inertia operator
${\mathcal I}\,:\, so(n)\to so(n)$. Then the Lagrangian of the free motion of
the body reads $l=\frac12\langle {\mathcal I}\omega,\omega\rangle$,
where now $\langle\cdot,\cdot\rangle$ denotes the Killing
metric on $so(n)$, $\langle X,Y\rangle=-\frac12\mbox{tr }(XY)$, $X,Y\in so(n)$.
For a ``physical'' rigid body, $\mathcal I\omega$ has the form
$I\omega+\omega I$, where $I$ is a symmetric $n\times n$ matrix called {\it mass
tensor} (see \cite{FeKo}). However, since we are interested mainly in
nonholonomic geodesic flows,
we shall consider other inertia operators as well.

Recall that in the three-dimensional case, the Suslov problem
describes the motion of a rigid body with the constraint:
the projection of the angular velocity to a vector fixed  in
the moving frame (for example $E_3$) is equal to zero \cite{Su, AKN}.
In other words, only infinitesimal rotations
in the planes span$(E_{1}, E_{2})$ and span$(E_{1}, E_{3})$ are allowed. Hence,
it is natural to define its $n$-dimensional analog
as follows: only infinitesimal rotations in the fixed 2-planes
spanned by $(E_{1}, E_{2}), \dots, (E_{1}, E_{n})$ (i.e., in the planes
containing
the vector $E_{1})$ are allowed.
Following \cite{FeKo}, one can relax these constraints by assuming that
the angular velocity matrix has the following structure
\begin{equation*}
\omega =
\begin{pmatrix}
0 & \cdots & \omega_{1r} & \cdots & \omega_{1n} \\
\vdots & \ddots & \vdots &  & \vdots \\
-\omega_{1r} & \cdots & 0 & \cdots & \omega_{rn} \\
\vdots &  & \vdots & \mathbf{O} &  \\
 - \omega_{1n} & \cdots & - \omega_{rn} &  &
\end{pmatrix},
\end{equation*}
where $\mathbf{O}$ is zero $(n-r)\times (n-r)$ matrix.
This implies the left--invariant constraints
\begin{equation}
\langle \omega, E_i\wedge E_j\rangle=0, \quad r+1\le i<j\le n.
\label{Suslov_constraints}
\end{equation}
 As a result, the Suslov problem is described by the EPS equations
\begin{equation}
\frac{d}{dt} \left({\mathcal I}\omega\right) =[\mathcal I\omega,\omega ]
+\sum_{r<p<q\le n}\lambda_{pq} \, E_p\wedge E_q ,
\label{Suslov_eq}
\end{equation}
together with Poisson equations (\ref{Poisson}).
Here the components of the vectors $e_1,\dots, e_n$ play the role of redundant
coordinates on $SO(n)$.

Various integrable cases of the Suslov problem with additional
potential fields and their multidimensional generalization are given in
\cite{KZ, Koz1, AKN, Ok} and \cite{Jo4}, respectively.

\subsection{Some Integrable Cases of EPS Equations}

\paragraph{EPS equations on symmetric pairs.} Let $\h$ be the subspace of the
algebra $\g$ spanned by $a^i,\; i=1,\dots,\rho$.

Consider the case when the tensor $\A$ preserves the orthogonal decomposition
$\g=\h+\D$, i.e., $\A=\A_\h+\A_\D$, where
$A_\h:\h\to \h$, $A_\D:\D\to \D$ are positive definite operators.
Then $\m=I(\D)=\D$, and we can write (\ref{EPS}) in the following way
\begin{equation}
\dot x=[x,\A_\D(x)]_\D, \qquad x\in \D ,
\label{EPS2}
\end{equation}
where $\xi_{\D}$ denotes the orthogonal projection
of $\xi\in\g$ to the subspace $\D$ (with respect to
$\langle\cdot,\cdot\rangle$).

Equation (\ref{EPS2}) preserve the standard measure on $\D$.
Also the Hamiltonian function
$H(x)=\frac12\langle x,\A_D(x)\rangle$ and the invariant
$F(x)=\langle x,x \rangle$ are always first integrals of the system.
Therefore, by the Jacobi theorem the equation (\ref{EPS2}) is always
integrable for $\dim\D\le 4$.

\begin{remark}{\rm
Note that, in general,  the invariant
$F(x)=\langle x,x \rangle$ is not the integral of (\ref{EPS}),
although it is always an integral of non-constrained system.
Namely, a first integral $f(x)$ of the Euler--Poincar\' e equations
$\dot x=[x,\A(x)]$ is the integral of (\ref{EPS}) if and only if the following
condition holds
\begin{equation}
\sum_i \lambda_i \langle \nabla f(x),a^i\rangle\vert_{x\in \mathfrak m}=0.
\label{integral}
\end{equation}
In our case $\nabla F(x)=2x$, $x\in\mathfrak m=\D$ is orthogonal to $\mathfrak
h$
and therefore the invariant $F(x)$ remains to be an integral.
}\end{remark}

\begin{example}\label{example}{\rm
Let $\mathfrak k$ be a subalgebra of $\g$
and $\mathfrak w$ the orthogonal complement of $\mathfrak k$.
Suppose that $(\g,\mathfrak k)$ is a {\it symmetric pair}, i.e.,
the following conditions are satisfied:
$$
[\mathfrak w,\mathfrak w]\subset \mathfrak k, \quad [\mathfrak k,\mathfrak
k]\subset \mathfrak k, \quad [\mathfrak k,\mathfrak w]\subset \mathfrak w.
$$
Then, in the special case $\mathfrak d=\mathfrak w$, we have $[\D,\D]_\D=0$.
Therefore all the solutions of equations (\ref{EPS2}) are constants.
As a result, the solution of the original system on $G$ (nonholonomic geodesic
lines of the metric $ds^2_\I$) is given by the motion along one-parameter subgroups,
$$
g(t)=g_0 \exp(t\xi), \qquad \xi\in\D.
$$
This simplest situation occurs for the multidimensional Suslov equations
(\ref{Suslov_eq}) with $r=n-1$ and $\mathcal I\omega=I\omega+\omega I$, where
$I=\mathrm{\diag}\,(I_1,\dots,I_n)$.
Then $\h=so(n-1)$, $(so(n),so(n-1))$ is a symmetric pair, and $\mathcal I$
preserves
the decomposition $so(n)=\D+so(n-1)$. Hence the solutions $\omega(t)$ are just
constants.
}\end{example}

Motivated by the above observation and by another integrable case
of the multidimensional Suslov problem (see below),
let us assume that there is a chain of subalgebras
$$
\mathfrak l\subset \mathfrak k\subset \g,
$$
where $(\mathfrak g,\mathfrak k)$ is a symmetric pair, and
consider the adjoint representation of $\mathfrak l$ onto the  linear space
$\mathfrak w$:
$\eta\in \mathfrak l \mapsto  [\eta,\cdot]\in End(\mathfrak w)$.
With respect to this representation, decompose $\mathfrak w$ into irreducible subspaces
$\mathfrak w=\mathfrak w_0+\mathfrak w_1+\dots+\mathfrak w_m$,
$\mathfrak w_0$ being the subspace with the trivial representation.
Next, assume that $\D$ has the form
\begin{equation}
\D=\mathfrak u+\D_0+\D_1+\dots+\D_g, \quad
\mathfrak u=\D\cap \mathfrak l,\quad \D_k=\D\cap \w_k.
\label{decomposition}
\end{equation}
with $\dim \mathfrak u \ge 1$, $g\le m$.
Suppose also that $A_\mathfrak u=s\cdot Id_{\mathfrak u}$, $s\in \mathbb{R}$
and that the operator $\A_\D$ preserves the decomposition (\ref{decomposition}),
that is, $\A_\D=A_\mathfrak u+A_0+\dots+A_g$.
Let $x_{k}$  denote the orthogonal projection of $x$ to $\D_k$, $k=0,\dots,g$.
Then the equation (\ref{EPS2}) reads
$$
\frac{d}{dt}(x_\mathfrak u + x_{0}+\dots+x_{g})=
[x_\mathfrak u + x_{0}+\dots+x_{g},s x_\mathfrak u +
A_0(x_{0})+\dots+A_g(x_{g})]_{\D}.
$$
In view of conditions $[\mathfrak u,\D_k]_{\D}\subset \D_k$,
$[\D_i,\D_j]_\D\subset \mathfrak u$, the above system splits into $g+2$ equations
\begin{align*}
\dot x_0 & =0,  \\
\dot x_\mathfrak u &=[x_{0}+\dots+x_{g},A_0(x_{0})
+\dots+A_g(x_{g})]_{\mathfrak u}, \\
\dot x_k & =[x_\mathfrak u,B_k(x_k)]_{\D_k}, \qquad k=1,\dots,g,
\end{align*}
where $B_k=A_k-s\cdot Id_{k}$.

Thus, apart from $H(x)=\frac12\langle \A_\D(x),x\rangle$ and
 $F(x)=\langle x,x \rangle$, the system (\ref{EPS2}) has a set of the first
integrals
given by the projection of $x$ to $\D_0$, $F_0(x)=x_{0}$  and the functions
$$
F_k(x)=\langle B_k (x_{k}),x_{k}\rangle,  \qquad k=1,\dots,g.
$$

In this case the following theorem holds (see \cite{Jo2}).
\begin{thm}
\begin{description}
\item{1).} If the operators $B_k$ are positive definite,
then invariant varieties
$$
M_{c}=\{ x\in D \mid  x_{0}=c_0, F_1(x)=c_1,\; \dots, F_n(x)=c_g, \;
F(x)=c_{g+1}\} ,
$$
$c_1,\dots, c_{g+1}$ being constants of motion,
are diffeomorphic  to the product of spheres
$
S^{\dim \D_1-1}\times \dots\times S^{\dim \D_g-1} \times S^{\dim \mathfrak u-1},
$
provided that $c_{g+1}$ satisfies inequality
\begin{equation}
c_{g+1}>\vert c_0 \vert^2 + \sum_{k=1}^g \frac{c_k}{b_k},
\quad b_k=\min_{\vert \xi_{k} \vert=1} \langle B_k(\xi_{k}),\xi_{k}\rangle \, .
\label{condition}
\end{equation}

\item{2)} If $\dim \D_k \le 2$, $k=1,\dots,g$, $\dim\mathfrak u =1$,  and
 all the constants $c_i$ are nonzero,
then $M_{c}$ is diffeomorphic to the disjoint union of two
$g$--dimensional tori with a quasi-periodic dynamic of the form
$\dot \varphi_i={\Omega_i}/{\Phi(\varphi)}$, $i=1,\dots,g$.
\end{description}
\end{thm}

\paragraph{The Fedorov--Kozlov case.}
The above construction applied to the symmetric pair
$(\mathfrak g,\mathfrak k)=(so(n),so(2)\times so(n-2))$
gives the Fedorov--Kozlov integrable case of the multidimensional Suslov problem \cite{FeKo}.
As above, let $\mathfrak w$ be the orthogonal complement of $\mathfrak k$:
$$
so(n) =
\left( \begin{array}{cc}   so(2) & \w \\
                     -\w^t & so(n-2) \\
               \end{array} \right).
$$
We take $\mathfrak u=\mathfrak l=so(2)=\Span\{E_1\wedge E_2\},\, \D=\mathfrak
u+\w$,
i.e., the constraint are given by relations (\ref{Suslov_constraints}) with
$r=2$. Then
$$
\D_1=\w_1=\Span\{E_1\wedge E_3, E_2\wedge E_3\},\; \dots, \;
\D_{n-2}=\w_{n-2}=\Span\{E_1\wedge E_n, E_2\wedge E_n\}.
$$

In the Suslov problem there is a natural choice of the inverse inertia operator
$\A=\I^{-1}$ which preserves the decomposition $\D=so(2)+\D_1+\dots+\D_{n-2}$.
Namely, we take the left-invariant metric on $so(n)$
determined by the kinetic energy of the multidimensional rigid body,
\begin{equation}
\label{physical}
\A: E_i \wedge E_j \mapsto \frac{1}{I_i+I_j} E_i\wedge E_j.
\end{equation}
If $I_1>I_2>I_3>\dots>I_n$, then, under condition (\ref{condition}),
the integrals $F_1,\dots,F_{n-2}$ are positive definite and the invariant
submanifolds
$$
\{F_1=c_1,\; \dots,\; F_{n-2}=c_{n-2}, \; F=c_{n-1}\}
$$
are union on two disjoint $(n-2)$--dimensional tori. Moreover, as shown in \cite{FeKo},
the motion on the tori is straight-line but not uniform and in appropriate angle
coordinates $\varphi_i$ it is described by equations
\begin{equation}
\dot\varphi_i=\frac{\Omega_i}{\omega_{12}(\varphi)},
\qquad
\Omega_i=\sqrt{\frac{(I_1-I_{i+2})(I_2-
I_{i+2})}{(I_1+I_{i+2})(I_2+I_{i+2})}},\quad
i=1,\dots,n-2 \, .
\label{FK}
\end{equation}

For this integrable case the reconstruction problem was
studied in \cite{Bl_Z}. As follows from (\ref{FK}),  if the trajectories
are periodic on one torus, they are periodic on the rest
of the tori. Then the trajectories $(g(t),\dot g(t))$
which correspond to the given periodic trajectory $\omega(t)$ are
quasi-periodic (e.g., see \cite{He}).

According to \cite{Bl_Z},  in the opposite case, if for some constants $c>0$ and
$\gamma>n-3$, the frequencies satisfy Diophantine conditions
$$
\vert l+\sqrt{-1}(k,\Omega)\vert \ge c/\vert k\vert^\gamma, \quad
l=0,1,2, \quad {\mathrm{for\;all}}\quad  k\in \mathbb Z^{n-2}
$$
and the value of the integrals $F$
(or $F$ and $F_1$) are dominant with respect to those of
other integrals ($c_i/c_{n-1}\sim \epsilon$, $i\ne n-1$ or
$c_i/c_1, c_i/c_{n-1} \sim \epsilon$, $i\ne 1,n-1$),
then the dynamics on the whole phase space can be
approximated by quasi--periodic dynamics on the
time interval of length $\sim \exp(1/\epsilon)$.

\paragraph{The Suslov problem on so(4).}
Now we concentrate on the integrable case when $\D$ is not an
eigenspace of $\A$. Let $\g=so(4)$. Then $\mathfrak
k=\Span\{E_1\wedge E_2,E_3\wedge E_4\}$ is a Cartan subalgebra. As
above, take the inertia operator in the form (\ref{physical}),
which implies $\A(\mathfrak k)=\mathfrak k$. Further, take $a=a_1
E_1\wedge E_2+a_2 E_3\wedge E_4$, $a_1a_2\ne 0$ and the constraint
$\langle a,\A(x)\rangle=0$.
 Then $[a,\A(a)]=0$ and the
Euler--Poincar\'e--Suslov equations
$$
\dot x=[x,\A(x)]+\lambda a
$$
preserve the standard measure on $\mathfrak m=\{x \mid \langle a,
\A(x)\rangle\}=0$ (see \cite{Koz2}). Next, one can always chose a
linear combination of quadratic invariants on $so(4)$,
$c_1I_1+c_2I_2$, such that the condition (\ref{integral}) holds.
Thus our system on five-dimensional space $\mathfrak m$ has the
integrals $H=\frac 12 \langle \A(x),x\rangle$ and
$F_1=c_1I_1+c_2I_2$. For the integrability one needs one more
independent integral. It can be taken in the form of a quadratic
function on the orthogonal complement of $\mathfrak k$.

This approach is a special case
of the method of construction of integrable EPS equations
on six-dimensional unimodular Lie algebras given in \cite{Jo1}.

\paragraph{Chains of subalgebras.}
Suppose there is a chain of subalgebras
$$
\g_0 \subset \g_1 \subset \dots \subset \g_n=\g.
$$
Let $\g_{i}=\g_{i-1}+\w_i$ be the corresponding orthogonal decompositions.
Then  $\g_i=\g_0+\w_1+\dots+\w_i$. Following \cite{Bo}, consider $\A$ of the form:
\begin{equation}
\A=\A_0+s_1\cdot Id_{\w_1}+\dots+s_n\cdot Id_{\w_n},
\qquad s_i>0,\quad i=1,\dots,n,
\label{chain_operator}
\end{equation}
where $\A_0$ is a symmetric positive operator  defined in the subalgebra $\g_0$.
Suppose that $\D$ has orthogonal decomposition
\begin{equation}
\D=\D_0+\D_1+\dots+\D_n,
\quad \D_k=\{\omega_k\in \w_k,\; \langle a_k^i,\omega_k \rangle =0, \;
i=1,\dots,\rho_k\}.
\label{chain_constraints}
\end{equation}
Then $\D_k$, $k>0$ are invariant subspaces of $\A$.
By $x_{k}$ denote the orthogonal projection of $x$ to $\D_k$,
$k>0$; and by $x_0$ denote the orthogonal projection to $\g_0$.

Now we can formulate the following theorem (see \cite{Jo2}).

\begin{thm}
\label{chain_th}
The  Euler--Poincar\' e--Suslov equations (\ref{EPS}),
with $D$ and operator $\A(x)$ of the form (\ref{chain_constraints}) and
(\ref{chain_operator}),
are equivalent to the  Euler--Poincar\' e--Suslov equations on the  Lie
subalgebra $\g_0$:
\begin{equation}
\dot x_0 = [x_0, \A_0(x_0)] + \sum_{i=1}^{\rho_0} \mu_i a^i_0,
\label{EPS3}
\end{equation}
$$
\langle \A_0(x_0),a^i_0 \rangle =0, \quad i=1,\dots, \rho_0,\\
$$
together with a chain of linear differential equations on the subspaces
$\D_k$:
\begin{equation}
\dot x_{k}=[x_{k}, \A_0(x_0)-s_kx_0+(s_1-s_k)x_{1}+\dots+(s_{k-1}-s_k)x_{k-
1}]_{k}.
\label{chain_eq}
\end{equation}
\end{thm}

If the Euler--Poincar\' e--Suslov equations (\ref{EPS3}) on $\g_0$ are
solvable, then the integration of original equations (\ref{EPS})
reduces to consequitive integration of the chain of
linear dynamical systems (\ref{chain_eq}) for $k>0$.
In the most simplest case the solutions of (\ref{EPS3}) are constants.
Then the components of the vector $x_1$ satisfy a system of linear equations with
constant coefficients, hence they are elementary functions of the time $t$.
This happens if $\A_0=Id_{\g_0}$ or if $\g_0$ is a commutative subalgebra.
In particular, if $\dim D_0=0$, then we have $x_0=0$.
In this case $\dot x_1=0$ and $x_2$ is given by elementary functions of $t$.

\subsection{Hamiltonian Flows}

In some cases, the nonholonomic geodesic flow (\ref{Hamilton}) on $\mathcal M$
can be obtained as a  restriction of a Hamiltonian flow on $T^*Q$ to
the invariant submanifold $\mathcal M$.
In most examples this happen when the Lagrange multipliers in  (\ref{Hamilton})
vanish, i.e., when $\mathcal M$ is an invariant submanifold of the unconstrained
geodesic flow.

There are also cases of nonzero  Lagrange multipliers.
This means that $\mathcal M$ is the invariant submanifold of some other
Hamiltonian system. In particular, in Example \ref{example} one can take a geodesic flow
of a bi-invariant metric.
Note that the Lagrange multipliers, in general, are different from zero
($\sum_{i=1}^{\rho} \lambda_i a^i=-[x,\A_\D(x)]$) and $\mathcal M$ is not an
invariant submanifold of the unconstrained geodesic flow of the left--invariant
metric $ds^2_\I$.

Below we concentrate on the first case (zero Lagrange multipliers).
Suppose that the orthogonal complement $\mathfrak h$ of $\D$ is the
Lie algebra $\mathfrak k$ of a Lie subgroup $\mathfrak H$ and that the operator
$\A$ also preserves orthogonal decomposition $\g=\mathfrak h+\D$, i.e.,
$\A=\A_\h+\A_\D$.
Then  $\m=\I(\D)=\D$ and EPS equations take the form (\ref{EPS2}).

Further, suppose that $\langle x,\A_\D(x)\rangle$ is an invariant
of the adjoint action of $\mathfrak H$ on $\D$,
$$
\langle [\mathfrak h, x],\A_\D(x)\rangle=0, \quad \mbox{or,
equivalently,} \quad [x,\A_\D(x)]_\mathfrak h=0, \quad x\in\D.
$$
Then one can easily check that $\D$ is the invariant subspace
of the Euler equations
\begin{equation}
\dot x=[x,\nabla H(x)]=[x,\A(x)], \quad x\in \g.
\label{Euler}
\end{equation}
Therefore the Lagrange multipliers vanish.
Moreover, one can consider the new Hamiltonian function
$H^*(x)=\frac12\langle x,\A_\D(x)\rangle$ on $\g$ and the
Euler equations
\begin{equation}
\dot x=[x,\nabla H^*(x)]=[x,\A_\D(x)], \quad x\in \g.
\label{sub-Riemannian}
\end{equation}

In both cases, the restriction of the systems to $\D$ coincides
with the Euler--Poincar\'e-Suslov equations (\ref{EPS2}). For
example, after projection to $\mathfrak h$ and $\D$ the system
(\ref{sub-Riemannian}) becomes
$$
\dot x_\D=[x_\D,\A_\D(x_\D)]+[x_\mathfrak k,\A_\D(x_\D)], \quad
\dot x_\mathfrak h=0.
$$

Let $h,\; h^*:T^*G\to {\mathbb R}$ be the functions
obtained by left translations from $H$ and $H^*$. While $h$ is
the Hamiltonian of the geodesic flow of the left-invariant metric $ds^2_\I$,
the function $h^*$ is degenerate in momenta and has another geometric meaning.

Suppose that $\D$  generates the Lie algebra $\g$ by commutations.
Then, by the Chow--Rashevski theorem,
any two points on $G$ can be joined by a piecewise smooth admissible curve $g(t)$.
Locally shortest admissible curves are called {\it sub-Riemannian geodesic
lines} of the sub-Riemannian metric
obtained by restriction of the given left-invariant metric $ds^2_\I$ to $D$.
The Hamiltonian flow of $h^*$ on $T^*G$ is a sub-Riemannian geodesic
flow. In other words, the projection of the flow to $G$ give us sub-Riemannian
geodesic lines (for more details see  \cite{Str, Ta}).
Such systems are also known as vaconomic systems \cite{AKN}.

We summarize previous considerations in the following proposition (see \cite{Jo2}).
 \begin{prop}
\label{subsystem}
Suppose that $H\vert_\D$ is an invariant
of $\mathfrak H$-adjoint action and that $\D$ generates the algebra $\g$
by commutations.
Then on the constrained submanifold $\mathcal M$
the following three different problems have the same flow:
the nonholonomic geodesic flow, the geodesic flow with
Hamiltonian $h$ and the sub-Riemannian geodesic flow
with Hamiltonian $h^*$.
\end{prop}

\paragraph{An example on the Lie group SU(n).}
Let us illustrate how the special case
of the construction given in the Theorem \ref{chain_th} produces a
nonholonomic geodesic flow with the above property.
Namely, consider the chain of subalgebras
$$
su(2)\subset  su(3)\subset \dots \subset su(n)
$$
given by the natural matrix embedding. Let $su(2+i)=su(2)+\w_i$ be
the orthogonal decompositions and let $\A$ has the form
\begin{equation}
\A= s_0 Id_{su(2)}+s_1\cdot Id_{\w_1}+\dots+s_{n-2}\cdot Id_{\w_{n-2}},
\quad s_i>0,\quad  i=0,\dots,n-2\, .
\label{chain_operator2}
\end{equation}
We take $\D$ to be the orthogonal complement to the Lie algebra of
the maximal torus $\mathbb T^{n-1}\subset SU(n)$ consisting of
diagonal matrices. Then the Hamiltonian $H=\frac12\langle
x,\mathcal A x\rangle$ will be an invariant of the adjoint action
of $\mathbb T^{n-1}$ on $su(n)$ (see, e.g., \cite{BoJo}) and $\D$
will generate $so(n)$ by commutations. Thus the system satisfies
the conditions of Proposition \ref{subsystem}.

Furthermore the system is integrable and can be considered as a
Chaplygin system as well. Namely, let $h$ be the corresponding
left invariant Hamiltonian function on $T^*SU(n)$. Since $H$ is
adjoint $\mathbb T^{n-1}$--invariant, we have that $h$ is also
{\it right} $\mathbb T^{n-1}$--invariant function. Thus, the group
$\mathbb T^{n-1}$ acts on Riemannian manifold
$(SU(n),ds^2_\mathcal I)$ by isometries. By submersion, the metric
$ds^2_\I$, induces the $SU(n)$-invariant metric $ds^2_{\I,sub}$ on
the flag manifold $F_n=SU(n)/\mathbb T^{n-1}$
\begin{equation}
\begin{array}{cccc}
\mathbb T^{n-1} & \longrightarrow & SU(n) &  \\
&  & \downarrow & \pi \\
&  &  F_n=SU(n)/ \mathbb T^{n-1} &
\end{array}.
\label{flag}
\end{equation}
Note that the horizontal spaces of the submersion
coinside with those of the distribution $D$.
In other words, we can also consider $(SU(n),ds^2,D,\mathbb T^{n-1})$
as an example of a Chaplygin system such that the right hand side
of (\ref{Chaplygin}) is equal to zero (see Remark \ref{submersions}).

The geodesic flow of the metric $ds^2_{\I,sub}$ on $F_n$ is
completely integrable (see \cite{BoJo}).
To describe the motion on the whole phase space $D$ one
must solve the reconstruction problem. Since the group $\mathbb T^{n-1}$
is Abelian, this can easily be done by quadratures (e.g., see \cite{MMR}).

\section{LR Systems}

\subsection{LR Systems as Generalized Chaplygin Systems}

Following \cite{VeVe1, VeVe2}, one defines an \textit{LR system}
on a compact Lie group $G$ as a nonholonomic Lagrangian system $(G,l,D)$
where $l$ is a left-invariant Lagrangian and $D$ is
a {\it right-invariant\/} distribution on $TG$.
As in LL systems, the Lagrangian is defined by a left-invariant metric
$ds^2_\I$, $l(g,\dot g)=\frac12\langle \I\omega,\omega\rangle$,
$\omega=g^{-1}\cdot \dot g$.

The distribution $D$ is determined by its restriction ${\D}$ to the Lie algebra,
$$
D_g=\D\cdot g= g\cdot (g^{-1}\cdot \D \cdot g) \subset T_g G.
$$
Let $\h=\Span\{a^1,\dots,a^{\rho}\}$ be the orthogonal
complement of $\D$ with respect to $\langle \cdot,\cdot \rangle$.
Then the  right-invariant constraints can be written as
\begin{gather}
\omega\in g^{-1}\cdot \D \cdot g, \quad {\mathrm{or}}\quad
\langle \omega,g^{-1}\cdot a^i \cdot g\rangle=0, \qquad  i=1,\dots,\rho,
\nonumber \\
\mbox{or, equivalently,  } \quad \langle \alpha^i,\A(x) \rangle=0, \qquad
\alpha^i=g^{-1}\cdot a^i \cdot g . \label{Rconstr}
\end{gather}
Equations (\ref{Hamilton}) in the left trivialization take the form
 \begin{eqnarray}
&&\dot x=[x,\A(x)]+\sum_{i=1}^{\rho} \lambda_i \alpha^i, \label{lr1}\\
&&\dot g=g\cdot \omega=g\cdot \A(x)\label{lr2}.
\end{eqnarray}
Here the Lagrange multipliers $\lambda_i$ are determinated by
differentiating the constraints.

The system (\ref{lr1}), (\ref{lr2}) is actually defined on the whole
phase space $TG$ and has first integrals
$$
f_i(g,x)=\langle \A(x),g^{-1}\cdot a^i \cdot g\rangle, \qquad i=1,\dots,\rho .
$$
Then the nonholonomic geodesic flow is just the restriction of (\ref{lr1}), (\ref{lr2})
onto the invariant submanifold  $\M=\{(g,x) \mid  f_i=0, \, i=1,\dots,\rho\}$.

Instead of  (\ref{lr1}), (\ref{lr2}),  one can consider the following closed
system on the direct product $\g^{1+\rho}$ in the variables
$\{x,\alpha^1,\dots,\alpha^\rho\}$,
\begin{eqnarray}
&&\dot x=[x,\A(x)]+\sum_{i=1}^{\rho} \lambda_i \alpha^i,
\label{lr3}\\
&&\dot \alpha^i=[\alpha^i,\A(x)],\qquad i=1,\dots,\rho, \label{lr4}
\end{eqnarray}
where the multipliers $\lambda_i$ are determined from the conditions
$\frac d{dt} \langle \alpha^i,\A(x)\rangle=0$.

Equations (\ref{lr4}) imply that $\alpha^i(t)$ belongs to
the adjoint orbit $O_G(\alpha^i(t_0))$.
Then, if $(x(t),\alpha^1(t),\dots,\alpha^\rho(t))$ is
a solution of (\ref{lr3}), (\ref{lr4})
and $g(t)$ is a solution of the kinematical equation (\ref{lr2})
(with appropriate initial conditions),
we conclude that $(g(t), x(t))$ is a solution of the system (\ref{lr1}), (\ref{lr2}).

One of remarkable properties of LR systems is the existence of an invariant
measure, which puts them rather close to Hamiltonian systems.
Veselov and Veselova \cite{VeVe2} proved that the system (\ref{lr3}),
(\ref{lr4}) has an invariant measure with density
\begin{equation}
\sqrt{\det\left(\langle \A(\alpha^i),\alpha^j\rangle\right)}. \label{density}
\end{equation}
This implies that the original system (\ref{lr1}), (\ref{lr2}) on $TG$
also has an invariant measure of the form $\mu(g)\cdot d\sigma$,
where $d\sigma$ is the canonical volume form on $TG$ and
$$
\mu(g)=\sqrt{\det\left(\langle \A(g^{-1}\cdot a^i\cdot g),g^{-1}\cdot a^j\cdot
g\rangle\right)}.
$$
In particular, our nonholonomic geodesic flow on $\M$ also has an invariant
measure described in the following way. Let $d\,\Sigma$ be a volume form on $\M$.
Then
\begin{equation}
d\sigma=\theta \; df_1 \wedge\cdots\wedge df_{\rho} \wedge d\,\Sigma, \quad
(g,x)\in \M
\label{decomp}
\end{equation}
for some positive function $\theta$.
Next, let ${\mathcal L}$ be the Lie derivative with respect to the flow
(\ref{lr1}), (\ref{lr2}).
Since the functions $f_s$ are first integrals, we have
${\mathcal L} d f_s=0$, $s=1,\dots,\rho$.  As a result, from the
condition ${\mathcal L}(\mu \; d\sigma)=0$ and (\ref{decomp}) we obtain
$df_1 \wedge\cdots\wedge df_{\rho} {\mathcal L}(\mu \theta \; d\Sigma)=0$.
Hence, the restriction of the flow onto $\M$ has the invariant measure
$\mu\theta \; d\Sigma$.

\paragraph{Reduction.}
Now, let the linear subspace $\mathfrak{h}$ be the Lie algebra of a subgroup
$H\subset G$.
Then the Lagrangian $l(g,\dot g)$  and the
right-invariant distribution  $D$  are also invariant with respect to the left
$H$-action.
Consider homogeneous space $Q=H \backslash G$ of cossets $\{Hg\}$. The
distribution $D$ can be seen as a principal connection of the principal bundle
\begin{equation*}
\begin{array}{cccc}
H & \longrightarrow & G &  \\
&  & \downarrow & \pi \\
&  & Q= H \backslash G &
\end{array}.
\end{equation*}

As a result, {\it the LR system $(G,l,D,H)$ can naturally be regarded
as a generalized Chaplygin system.}
In order to write the reduced system on $Q$ in a simple form,
we identify $\mathfrak{g}$ and $\mathfrak{g}^*$ by the $Ad_G$--invariant scalar
product $\langle \cdot,\cdot \rangle$, and the spaces $TQ$, $T^*Q$ by the {\it
normal\/} metric, induced by the bi-invariant metric on $G$. Next, consider the
moment mappings:
\begin{equation*}
\phi: TG \cong T^*G\to \mathfrak{g}, \quad \Phi: TQ\cong T^*Q \to
\mathfrak{g},
\end{equation*}
of the natural \textit{right} actions of $G$ on $T^*G$ and $T^*Q$, respectively.
We have $\phi(\dot g)=\omega=g^{-1}\cdot \dot g$ and
the map $\Phi$ can be considered as a restriction of $\phi$ to $D$.

The {reduced Lagrangian} is, by definition, the constrained Lagrangian
$$ l_c(g,\dot g)=\frac12\langle \pr_{g^{-1}\mathfrak d
g} \mathcal I(\phi(g,\dot g)),\phi(g,\dot g)\rangle\, ,
$$
considered on the orbit space $H\backslash D \cong T(H\backslash G)$.
It follows that the reduced Lagrangian is simply given by
\begin{equation*}
L(q,\dot q)=\frac12 \langle {\mathcal I} \Phi(q,\dot q), \Phi(q,\dot q) \rangle,
\end{equation*}
where $q=\pi(g)$ are local coordinates on $Q$ (which may be
redundant). This is a Lagrangian of the geodesic flow of metric
which we shall denote by $ds^2_{{\mathcal I},D}$.

By using equations (\ref{Chaplygin}) one can prove the following
proposition (see \cite{FeKo}), which is a special case of the
general nonholonomic reduction procedure described in \cite{Koi,BKMM}.

\begin{prop}
The reduced Lagrange--d'Alambert equation describing the motion of the LR system
$(G,l,D)$ has the form
\begin{equation}
\left( \frac{\partial L}{\partial q} - \frac{d}{dt}
\frac{\partial L}{\partial \dot q}, \xi\right) =
\langle {\mathcal I}\Phi(q,\dot q), \pr_{g^{-1}\mathfrak{h} g}
 [\Phi(q,\dot q),\Phi(q, \xi)] \rangle,
\label{reduced}
\end{equation}
for all virtual displacements $\xi\in T_q Q$, where
$\pr_{g^{-1}\mathfrak{h} g}: \mathfrak{g}\to g^{-1}\mathfrak{h} g$
is the orthogonal projection, and $q=\pi(g)$.
\end{prop}

In addition, it appears that the reduced LR system (\ref{reduced}) also
possesses an invariant measure
(note that a generic Chaplygin system does not have this property, see
\cite{CCLM}).
Namely, the following general statement holds (e.g., see \cite{FeJo}).

\begin{lem}
Suppose there is a compact group $\mathfrak G$
acting freely on a manifold $N$ with local coordinates z and
there is a $\mathfrak G$--invariant dynamical system $\dot z=Z(z)$ on $N$.
If this system has an
invariant measure (which is not necessary $\mathfrak G$-invariant), then the
reduced
system on the quotient manifold $N/\mathfrak G$ also has an invariant measure.
\end{lem}

\subsection{Veselova Problem, an Integrable Geodesic Flow
on the Sphere and the Neumann Problem}

\paragraph{Veselova problem.}
The most descriptive illustration of an LR system is the
\textit{Veselova problem} on the  motion of a rigid body
about a fixed point under the action of nonholonomic constraint
\begin{equation}
\label{classical}
(\Omega,\gamma )=0,
\end{equation}
where $\Omega\in {\mathbb R}^3$ is the angular velocity vector,
$\gamma\in {\mathbb R}^3$ is a unit vector, which is fixed
in a space frame, and $(\, ,\, )$ denotes the scalar product in ${\mathbb R}^3$
\cite{VeVe1}.
Geometrically this condition means that the projection of the angular velocity
of the body to a fixed vector must equal zero.

The equations of motion in the moving frame have the form
\begin{eqnarray}
{\mathcal I}\dot \Omega ={\mathcal I}\Omega \times \Omega
+ \lambda \gamma,  \quad \dot \gamma = \gamma \times \Omega,
\label{3.1}
\end{eqnarray}
where $\mathcal I$ is the inertia tensor of the rigid body, $\times$ denotes the
vector product in ${\mathbb R}^3$, and  $\lambda$ is a Lagrange
multiplier chosen such that $\Omega(t)$ satisfies the above constraint,
\begin{equation} \label{la}
\lambda=- \frac {({\mathcal I}\Omega \times \Omega,
{\mathcal I}^{-1}\gamma)}{({\mathcal I}^{-1}\gamma,\gamma ) }\, .
\end{equation}
The Veselova system (\ref{classical}), (\ref{3.1}) is an LR system on the
Lie group $SO(3)$, which is
the configuration space of the rigid body motion. After identification of Lie
algebras $(\mathbb{R}^3,\times)$ and $(so(3),[\cdot,\cdot])$, the operator
$\mathcal I$ induces the left-invariant metric $ds^2_I$. The angular velocity
correspond to $\Omega=g^{-1}\dot g$, the velocity in the left
trivialization $TSO(3)\cong SO(3)\times so(3)$.
The  vector fixed in the
space corresponds to the right-invariant vector field $\gamma_g=g\cdot
(g^{-1}\cdot a \cdot g)\in T_g SO(3)$, $a\in so(3)$, and the
nonholonomic constraint (\ref{classical}) has the form
$\langle g^{-1} \cdot a \cdot g, \Omega\rangle=0$.
Once can check that the closed system (\ref{3.1}), (\ref{la}) has invariant
measure with density $(\gamma, {\cal I}^{-1}\gamma)$, as predicted by formula
(\ref{density}). Note that
integrable potential perturbations of the Veselova system can be found in
\cite{VeVe1, FeJo}.

\paragraph{Multidimensional Veselova problem.}
Now we proceed to a $n$-dimensional generalization of the Veselova system,
describing the motion on the Lie group $SO(n)$
with certain right-invariant nonholonomic constraints.

Let, as above, $e_1,\dots, e_n$ be unit vectors that form a fixed orthogonal
frame
in the space ${\mathbb R}^n$. Then, similarly to the generalized Suslov
problem in Section 2, we define $n$-dimensional analog of (\ref{classical})
as follows: only infinitesimal rotations in the fixed 2-planes
spanned by $(e_{1}, e_{2}), \dots, (e_{1}, e_{n})$ are allowed.
This implies the constraints
\begin{equation}
\langle \omega, e_i\wedge e_j\rangle=0, \qquad 2\le i<j\le n.
\label{Veselova_constraints}
\end{equation}
Equivalently, consider the right-invariant distribution $D$ on $TSO(n)$
whose restriction to the algebra $so(n)$ is given by
$\mathfrak{d}=\Span\{E_j \wedge E_k, \; k=1,\dots, r, \; j=1,\dots, n\}$,
where $E_i \wedge E_j$ form the basis in $so(n)$.
Since $e_i \wedge e_j=g^{-1} \cdot E_i \wedge E_j \cdot g$, we have that
constraints are
$
\omega \in {\cal D}= g^{-1}\cdot \mathfrak{d} \cdot g=\Span\{e_1 \wedge e_i, \;
2\le i \le n\}.
$

\begin{remark}{\rm
As for the multidimensional Suslov problem, the constraints
(\ref{Veselova_constraints})
can be relaxed. However, in this case,
the existence of the integrable LR system is still not known.
That is why we keep using the above constraints (see Theorem \ref{main}).
}\end{remark}

The LR system can be described by the
Euler--Poincar\'e  equations (\ref{lr1}, \ref{lr2}, \ref{lr3})
on the space $so(n)\times SO(n)$ with indefinite multipliers $\lambda_{pq}$,
\begin{align}
&\frac{d}{dt}\left({\mathcal I}\omega\right) =[{\mathcal I}\omega ,\omega]
+\sum_{2\le p<q\le n}\lambda_{pq} \, e_p\wedge e_q , \nonumber \\
&\dot{e}_{i}+\omega e_{i} =0 , \qquad i=1,\dots, n .
\label{Veselova_eq}
\end{align}
Here, as above, the components of $e_1,\dots, e_n$ play the role of
redundant coordinates on $SO(n)$.

\paragraph{Reduction.}
The orthogonal complement $\mathfrak{h}$ of $\mathfrak{d}$ is
a Lie algebra, namely
$$
\mathfrak{h}=\Span\{E_p \wedge E_q, \; 2\le p<q \le n\}\cong so(n-1).
$$
Therefore, the Veselova system can be treated as a
generalized Chaplygin system on the principal bundle
\begin{equation}\begin{array}{cccc}
SO(n-1) & \longrightarrow & SO(n) &  \\
&  & \downarrow & \pi \\
&  &  S^{n-1}=SO(n-1)\backslash SO(n) &
\end{array},
\end{equation}
where $S^{n-1}$ is the $n$-dimensional sphere, realized
as  the unit sphere in $\mathbb{\ R}^n$,
\begin{equation*}
S^{n-1}=\{q\in\mathbb{\ R}^{n-1}, \; q_1^2+\cdots+q_n^2=1\},
\end{equation*}
where we set $q=e_1$. The moment map is then
$
\omega=\Phi(q,\dot q)=q\wedge \dot q.
$
Thus, for solution $e_1(t)$, $\omega(t)=e_1(t) \wedge \dot e_1(t)$ of
(\ref{Veselova_eq}), $q(t)=e_1(t)$ is a motion of a reduced system on
the sphere $S^{n-1}$.

\paragraph{The invariant measure.}
It appears that for some special inertia tensors, many of the calculations
takes an especially simple form.
Suppose that the operator ${\mathcal I}$ is defined by a diagonal matrix
$A=\diag(A_1,\dots,A_n)$
in the following way
\begin{equation}
{\mathcal I} (E_i\wedge E_j)=\frac{A_iA_j}{\det A} E_i\wedge E_j  .
\label{inertia_tensor}
\end{equation}
Notice that for $n=3$ this corresponds to the well known three-dimensional
vector formula
$I(x\times y)=(\det A)^{-1}\, Ax \times Ay$, $A=I^{-1}$.

Under the condition  (\ref{inertia_tensor}) the reduced Lagrangian $L(q,\dot q)$
and the right hand side of the Lagrange-d'Alambert equation
(\ref{reduced}) take the form
\begin{align}
L & = \frac 1{2\det A} [(A\dot q,\dot q)(Aq,q)- (Aq,\dot q)^2 ] \, ,\label{redL} \\
 \langle {\cal I} \Phi(q,\dot q), & \pr_{g^{-1}\mathfrak{h} g}
 [\Phi(q,\dot q),\Phi(q, \xi)] \rangle  = \frac 1{\det A} \langle A q \wedge
A\dot q,
\pr_{g^{-1}\mathfrak{h} g}\xi \wedge \dot q \rangle  \nonumber \\
 & = \frac 1{\det A} (\dot q, A \dot q) (Aq, \xi) -  \frac 1{\det A} (\dot q, A q)
(A\dot q, \xi) = \Psi(q,\dot q,\xi)\, .
\label{Pi}
\end{align}
Here we used relation $\pr_{g^{-1}\mathfrak{h} g}\xi \wedge \dot q = \xi
\wedge\dot q$
for any admissible vector $\xi=(\xi_1,\dots,\xi_n)^T \in T_q S^{n-1}$.
Below we shall keep using the redundant coordinates $q_i$ and velocities
$\dot q_i$, in which the Lagrange equations have the form
 \begin{gather}
\frac{\partial L}{\partial q_i} - \frac{d}{dt}\frac{\partial
L}{\partial \dot q_i} =\pi_i +\Lambda q_i \, , \qquad i=1,\dots, n,
 \label{sphere_eq} \\
\pi_i= \frac{\partial\Psi}{\partial \xi_i}
=\frac 1{\det A}  (\dot q, A \dot q) A_i q_i - \frac 1{\det A}  (\dot q, A q)
A_i \dot q_i, \nonumber
\end{gather}
where $\Lambda$ is a Lagrange multiplier.

Now we want to represent
the reduced LR system on $T^*S^{n-1}$ as a restriction of a system on
the Euclidean space ${\mathbb R}^{2n}=\{q, p\}$.
Note that $L(q,\dot q)$ is degenerate in the redundant velocities $\dot q$,
hence they cannot be expressed uniquely in terms of the redundant moments
\begin{equation}
\label{moments}
p_i= \frac{\partial L}{\partial\dot q_i}
\equiv \frac 1{\det A} (q, A q) A_i \dot q_i
-\frac 1{\det A} (\dot q, A q) A_i q_i .
\end{equation}
In this case one can apply the Dirac formalism for Hamiltonian systems with
constraints in the phase
space  (see, e.g., \cite{Dirac, AKN, Moser_var}).  Namely, from (\ref{moments})
we find that $(q,p)=0$,
hence the cotangent bundle $T^*S^{n-1}$ is realized as a subvariety of
${\mathbb R}^{2n}=(q,p)$ defined by constraints
$$
\phi_1\equiv (q,q)=1, \quad \phi_2\equiv (q,p)=0.
$$
Under these conditions, relations (\ref{moments}) can be uniquely inverted to
yield
\begin{equation}
\label{dot_q}
\dot q=\frac {\det A}{(q,Aq)} \left[ A^{-1}p - (p,A^{-1}q) q\right]\, .
\end{equation}
On the other hand, we note that $\partial L/\partial q_i=\pi_i$. Then, from
(\ref{sphere_eq})
we obtain $\dot p=-\Lambda q$ and, from the condition $(\dot q,p)+(q,\dot p)=0$,
\begin{equation} \label{dot_p}
\dot p=-\Lambda q, \qquad \Lambda=\det A \frac {(p,A^{-1}p) - (p,q) (q,A^{-1}p)}
{(q,Aq)}\,  .
\end{equation}

The system (\ref{dot_q}), (\ref{dot_p}) on $T^* S^{n-1}$ coincides with the
restriction of
the following system on  ${\mathbb R}^{2n}=\{q, p\}$
\begin{gather*}
\dot q_i=\{q_i, \hat H \}_*, \quad \dot p_i=\{p_i, \hat H \}_*  -\hat \pi_i \, , \\
\hat\pi_i(q,p) =\pi_i (q,\dot q(q,p)), \quad \hat H=\frac 12 \det A \frac{ (p,
A^{-1}p)}{(q,Aq)}\,  ,
\end{gather*}
which is quasi-Hamiltonian with respect to the following Dirac bracket on ${\mathbb R}^{2n}$
$$
\{F, G\}_* =\{F, G\}
+\frac{ \{F,\phi_1 \}\{G,\phi_2\}- \{F,\phi_2\} \{G,\phi_1\} }
{ \{\phi_1, \phi_2 \} },
$$
$\{\cdot ,\cdot\}$ being the standard Poisson bracket on ${\mathbb R}^{2n}$.
This system has explicit vector form
\begin{equation}
\label{extended}
\begin{aligned}
\dot q & = \frac {\det A}{(q,Aq)} \left[ A^{-1}p - \frac{(p,A^{-1}q)}{(q,q)}
q\right]\, ,  \\
\dot p & = - \det A \frac {(p,A^{-1}p) (q,q)- (p,q) (q,A^{-1}p)}
{(q,Aq)(q,q)^2}\,q\,  .
\end{aligned}
\end{equation}
The bracket $\{\cdot ,\cdot\}_*$ is degenerate and possesses Casimir functions
$\phi_1, \phi_2$ specified above.

Now, we can find the explicit form of the invariant measure of the reduced
system. From (\ref{extended}) we find
$$
\sum_{i=1}^n \left( \frac{\partial \dot q_i}{\partial q_i}
+\frac{\partial \dot p_i}{\partial p_i}\right)
=- (n-2) \frac{\det A\, (p, A^{-1}q)}{(q,q)\, (q,Aq)},
$$
which, in view of (\ref{moments}), takes the form $(n-2) (q, A \dot q)/(q,Aq)$.
Hence the extended system  (\ref{extended}) possesses an invariant measure
$$
{\cal J}=(Aq,q)^{-(n-2)/2}\, dp_1\wedge dq_1\wedge\cdots\wedge dp_n\wedge dq_n.
$$
Next, at points of $T^* S^{n-1}$, the standard volume form in ${\mathbb R}^{2n}$
can be represented as
$$
dp_1\wedge dq_1\wedge\cdots\wedge dp_n\wedge dq_n
= {\bf w}^{n-1}\wedge d\Phi_1 \wedge d\Phi_2,
$$
where ${\bf w}$ is the restriction of the standard
symplectic form $dp_1\wedge dq_1+\cdots+dp_n\wedge dq_n$ onto $T^* S^{n-1}$
and $\Phi_1,\Phi_2$ are certain functions of the Casimir functions
$\phi_1, \phi_2$. Since the latter are invariants of the vector field
$V(p,q)$ given by (\ref{extended}), the Lie derivatives
${\cal L}_V d\Phi_1, {\cal L}_V d\Phi_2$  equal zero. Then,
since ${\cal L}_V {\cal J}=0$, we conclude that on  $T^* S^{n-1}$,
$$
{\cal L}_V\, [ (Aq,q)^{-(n-2)/2}\, {\bf w}^{n-1} ] =0.
$$

As a result, we arrive at the following theorem.
\begin{thm}
\label{redLR}
The reduced LR system (\ref{dot_q}, \ref{dot_p}) on $T^*S^{n-1}$ possesses an
invariant measure
$$
f(q)=(Aq,q)^{-(n-2)/2}\, \sigma, \qquad \sigma={\bf w}^{n-1}
$$
where $\sigma$ is the canonical volume $2(n-1)$-form on $T^*S^{n-1}$.
\end{thm}

\paragraph{Chaplygin reducing multiplier.}
As follows from Theorem \ref{redLR}, item 1) of Theorem
\ref{Th_reduce}, and the fact that the dimension of the reduced
configuration manifold equals $n-1$, if our reduced LR system on
$T^* S^{n-1}$ were transformable to a Hamiltonian form by a time
reparameterization, then the corresponding reducing multiplier
$\N$ should be proportional to $1/\sqrt{(q, Aq)}$.

Although Chaplygin's reducibility theorem does not admit a
straightforward multidimensional generalization, i.e.,  item 1) of
Theorem \ref{Th_reduce} cannot be inverted, remarkably, for our
reduced LR system on $T^* S^{n-1}$ the inverse statement becomes
applicable (see \cite{FeJo}).

\begin{thm}
\label{main}
\begin{description}
\item{1).} Under the time substitution  $d\tau =\sqrt{\det A/
(Aq,q) }\, dt$ and an appropriate change of momenta, the reduced
LR system (\ref{sphere_eq}) or (\ref{dot_q}), (\ref{dot_p})
becomes a Hamiltonian system describing a geodesic flow on
$S^{n-1}$ with the Lagrangian
\begin{equation} \label{L^*}
L^{\ast }(q,d q/d\tau )=\frac{1}{2} ( q, Aq)^{-1} \left[ \bigg(A
\frac{d\, q}{d\tau}, \frac{d\, q}{d\tau} \bigg)(Aq,q) - \bigg(Aq,
\frac{d\, q}{d\tau}\bigg)^2 \right]\, .
\end{equation}

 \item{2).} For $A_1<A_2<\dots<A_n$ the latter system is algebraic
completely integrable for any dimension $n$. In the spheroconic
coordinates $\lambda_{1}, \dots,\lambda _{n-1}$ on $S^{n-1}$\ such
that
\begin{equation} \label{S_n}
q_{i}^{2}=\frac{\left( I_{i}-\lambda _{1}\right) \cdots
\left( I_{i}-\lambda_{n-1}\right) }{\prod_{j\ne i} \left( I_{i} -I_{j}\right) },
\qquad I_i=A^{-1}_i
\end{equation}
the Lagrangian $L^* (q, dq/d\tau )$ takes the St\"ackel form
\begin{equation*}
L^{\ast }  =\frac{1}{8}\sum\limits_{k=1}^{n-1}
\frac{\prod_{s\neq k}\left( \lambda _{k}-\lambda _{s}\right) }
{\left( \lambda_{k}-I_{1}\right)
\cdots \left( \lambda _{k}-I_{n}\right) \lambda _{k}}\left( \frac{d}{d\tau }
\lambda _{k}\right) ^{2},
\end{equation*}
and the evolution of $\lambda_k$
is described by the Abel--Jacobi quadratures
\begin{gather} \label{quad_N}
\frac{\lambda _{1}^{k-1}d\lambda _{1}}{2\sqrt{R\left( \lambda _{1}\right) }}
+\cdots +\frac{\lambda _{n-1}^{k-1}d\lambda _{n-1} }{2\sqrt{R\left( \lambda
_{n-1}\right) }}=\delta _{k,n-1}\, \sqrt{2h}\, d\tau,  \\
k=1, \cdots ,n-1, \nonumber
\end{gather}
where
\begin{equation} \label{Rad}
R(\lambda) =- ( \lambda -I_{1}) \cdots (\lambda-I_{n})
\lambda (\lambda-c_{2})\cdots (\lambda-c_{n-1}),
\end{equation}
$h=L^*$ being the energy constant and $c_{2},\cdots,c_{n-1}$ being other
constants of motion
(we set $c_1=0$).
For generic values of these constants the corresponding invariant manifolds are
$(n-1)$-dimensional
tori.
\end{description}
\end{thm}

The item 1) of Theorem \ref{main} is based on the relation between
the reduced LR system to the celebrated Neumann system (see
Theorem \ref{LR->Nn} below).

Namely,  consider the iso-energy submanifold $\mathcal E_h=\{L(q,\dot
q)=h\}\subset TS^{n-1}$ of the reduced
Veselova system (\ref{sphere_eq}) and introduce another new time $\tau_1$ by
formula
\begin{equation}
d\tau_1=\sqrt{\det A \frac{2h} {(Aq,q)}}\, dt.
\label{tau_1}
\end{equation}

\begin{thm}  \label{LR->Nn}
Under the time substitution (\ref{tau_1}), the solutions $q(t)$ of
the reduced multidimensional Veselova system on $S^{n-1}$ lying on
the $\mathcal E_h$ transforms to the solution of the integrable
Neumann problem with the potential $U(q)=\frac12 (A^{-1} q,q)$,
\begin{equation}
\frac{d^2}{d\tau_1^2} q= -\frac{1}{A} q + \lambda q
\label{Neumann}
\end{equation}
corresponding to zero value of the integral
\begin{equation}
F_0 =\left(
A\frac{dq}{d\tau_1},\frac{dq}{d\tau_1}\right ) (Aq,q)-
\left ( Aq, \frac{dq}{d\tau_1}\right )^2-(A q,q)
\label{F_0}\end{equation}
and vise versa.
\end{thm}

For $n=3$, Theorem \ref{LR->Nn} is proved by Veselov and Veselova \cite{VeVe2}.
The proof for arbitrary dimensions is given in \cite{FeJo}.

\subsection{Reconstructed Motion on $D$}
Now we consider the integrability of the original (unreduced) LR system on the
right-invariant distribution $D \subset TSO(n)$, which is specified by constraints
(\ref{Veselova_constraints})
and the left-invariant metric given by  (\ref{inertia_tensor}).
The relation between the reduced LR system and the Neumann system
described by Theorem \ref{LR->Nn} appears to be useful to reconstruct the motion
on $D$ exactly.
For this purpose we also shall make use of the correspondence between the
Neumann system and the geodesic flow on a quadric (see Kn\"orrer \cite{Knorr1}).
Namely, consider a family of $(n-1)$-dimensional confocal quadrics in ${\mathbb
R}^n$,
\begin{equation}
Q(\alpha)= \bigg\{ \frac{X^2_1} {\alpha -A_1} +\cdots+\frac{X^2_n}{\alpha -
A_n}=-1 \bigg\}\, ,
\quad \alpha\in {\mathbb R}.
\end{equation}

\begin{thm}
\label{Kno} \textup{(\cite{Knorr1})}.
Let $ X(s)$ be a geodesic on the quadric $Q(0)$, $s$ being a
natural parameter. Then under the change of time
\begin{equation}
ds =\sqrt{\frac {(dX/ds, A^{-1} dX/ds)}{(X, A^{-2} X) }}\,d\tau_1
\label{knorrer}
\end{equation}
the unit normal vector $q(\tau_1)= A^{-1} X /| A^{-1} X|$ is a
solution to the Neumann system (\ref{Neumann}) corresponding to zero
value of the integral (\ref{F_0}) and vise versa.
\end{thm}

It is well known that the problem of geodesics on a  quadric $Q(0)$ is
completely integrable, and qualitative behavior of the geodesics is described
by the remarkable {\it Chasles theorem\/} (see e.g., \cite{Knorr1, Moser_var}):
the tangent line
$
\ell_s=\{X(s)+\sigma\, dX/ds \mid \sigma \in{\mathbb R} \}
$
of a geodesic $X(s)$ on $Q(0)$ is also tangent to a fixed set of confocal
quadrics $Q(\alpha_2),\dots, Q(\alpha_{n-1})\subset {\mathbb R}^n$, where
$\alpha_2,\dots, \alpha_{n-1}$ are parameters
playing the role of constants of motion (we set $\alpha_1=0$). Now let
${\mathfrak n}_k$ be
the normal vector of the quadric $Q(\alpha_k)$ at the touching point
${\mathfrak p}_k=\ell \cap Q(\alpha_k)$. Then another classical theorem of
geometry says that
the normal vectors ${\mathfrak n}_1,\dots, {\mathfrak n}_{n-1}$, together with
the unit tangent
vector $\gamma=dX/ds$, form an orthogonal basis in $\mathbb{R}^{n}$.

On the other hand, in \cite{Moser_var}, Moser made the following observation.
\begin{prop}
\label{Moser1}
\begin{description} \item{1).} Let $x$ be the position vector of a point on the
line $\ell_s$, which is tangent to geodesic $X(s)$. Then in the new
parameterization $s_1$ such that $ds=- (X, A^{-2} X) \,ds_1$ the
evolution of the line  is described by the Lax equations in $n\times n$ matrix form
\begin{gather}
\frac d{ds_1} {\mathcal L}= [{\mathcal L}, {\mathcal B}], \qquad
{\mathcal L} = \Pi_ \gamma (A- x\otimes x)\Pi_\gamma, \\
{\mathcal B} = A^{-1} x\otimes A^{-1} \gamma - A^{-1} \gamma\otimes A^{-1} x\, ,
\label{calB}
\end{gather}
where $\Pi_\gamma=Id- (\gamma,\gamma)^{-1}  \gamma \otimes \gamma$ is the
projection onto
the orthogonal complement of $\gamma$ in ${\mathbb R}^n$.
\item{2).} The conserved eigenvalues of ${\mathcal L}$ are given by the
parameters
$\alpha_1=0, \alpha_2,\dots, \alpha_{n-1}$ of the confocal quadrics and by an
extra zero. The corresponding eigenvectors are parallel to the normal vectors
${\mathfrak n}_1=q,\dots, {\mathfrak n}_{n-1}$, and to $\gamma$.
\end{description}
\end{prop}

Now we are ready to describe generic solutions of the original LR
system on $D\subset T SO(n)$. Let  $q(\tau_1)$ be the solution of
the Neumann system (\ref{Neumann}) with $F_0 (q,q^{\prime})=0$, which
is associated to a solution $(q(t), p(t))$ of the reduced LR
system as described by Theorem \ref{LR->Nn}. Let
\begin{equation}
\label{Gauss}
X=(q,Aq)^{-1/2} A q(s), \,{\mathfrak n}_1=q(s),\dots, {\mathfrak
n}_{n-1}(s), \, \gamma(s)=\frac{dX}{ds}
\end{equation}
be the corresponding geodesic on $Q(0)$ in the new parameterization
$s$ given by (\ref{knorrer}) and the unit eigenvectors of $\cal
L$. Also, according to (\ref{tau_1}) and (\ref{knorrer}) we can
treat $s$ as a known functions of the original time $t$. Then we
have the following reconstruction theorem (see \cite{FeJo}).

\begin{thm} \label{recon0}
A solution $(g(t), \dot g(t))$ of the original LR system on the
distribution $D$ is given by the momentum map
$\omega(t)=q\wedge\dot q$ and the orthogonal frame formed by the
unit vectors
$$
e_1=q(t),\; e_2= {\mathfrak n}_2(t),\; \dots,\; e_{n-1}={\mathfrak n}_{n-
1}(t),\quad e_n=\gamma(t).
$$
The other solutions $(g(t), \dot g(t))$ that are projected onto
the same trajectory $(q(t), p(t))$  have the same $\omega, e_1$,
while the rest of the frame is obtained by the orthogonal transformations,
\begin{equation} \label{orth}
(e_2 (t) \cdots e_n(t) )
=({\mathfrak n}_2(t) \cdots {\mathfrak n}_{n-1}(t)\,  \gamma(t) )\,{\mathfrak R},
\end{equation}
where the constant matrix ${\mathfrak R}$ ranges over the group $SO(n-1)$.
\end{thm}

Thus, from Theorems \ref{recon0}, \ref{LR->Nn} and the integrability
properties of the Neumann system on $T^* S^{n-1}$ we conclude that
the phase space $D\subset T\,SO(n)$ of the multidimensional
Veselova LR system with the left-invariant metric defined by
(\ref{inertia_tensor}) is almost everywhere foliated by $(n-1)$-dimensional
invariant tori, on which the motion is straight-line but not uniform.

\subsection{Veselova Problem with Integrable Potentials and the Maupertuis Principle}

\paragraph{The Maupertuis principle.}
Consider a natural mechanical system on a compact Riemannian manifold $(Q,ds^2)$
with Hamiltonian $h(q,p)=\frac12\sum g^{ij}(q)p_ip_j+v(q)$, where
$g^{ij}$ is the inverse of the metric tensor and $v(q)$ is a smooth potential on
$Q$. Let
By the classical {\it Maupertuis principle}, the integral trajectories of the
Hamiltonian vector field $X_h$ with  $h(q,p)=c>\max v(q)$ coincide (up to a reparametrization)
with the trajectories of another vector field
$X_{h^J}$ with Hamiltonian
$$
h^J(q,p)=\frac12\sum\frac{g^{ij}(q)}{c-V(q)}p_i p_j
$$
on the fixed iso-energy level $\mathcal E_{c}=\{h(q,p)=c\}=\{h^J(q,p)=1\}$.
Namely, on $\mathcal E_{c}$ we have $dh=(c-v)dh^J$ (see \cite{AKN}).
The Hamiltonian flow of $h^J$ is the geodesic flow of the Jacobi metric
$ds^2_J=(c-v(q))ds^2$,which is conformally equivalent to the original metric $ds^2$.

The Maupertuis principle can  naturally be formulated for nonholonomic systems as well.
Suppose the distribution $D$ is locally defined by
$\rho=n-k$ independent 1-forms $\alpha^i$.
Then the equations of the nonholonomic systems with Hamiltonians
$h$ and $h^J$ subjected to the constraints $\dot q\in D_q$ are given by
\begin{eqnarray}
&&\dot p_i=-\frac{\partial h}{\partial q_i}+
\sum_{i=1}^{\rho} \lambda_j \alpha^j(q)_i\, ,
\quad \dot q_i=\frac{\partial h}{\partial p_i}\, , \qquad i=1,\dots,n,\label{Hamilton2}\\
&&\dot p_i=-\frac{\partial h^J}{\partial q_i}+
\sum_{i=1}^{\rho} \mu_j \alpha^j(q)_i\, ,
\quad \dot q_i=\frac{\partial h^J}{\partial p_i}\, , \qquad i=1,\dots,n.\label{Hamilton3}
\end{eqnarray}
On the iso-energy level $\mathcal E_{c}$, the vector fields (\ref{Hamilton2}) and
(\ref{Hamilton3}) are proportional
and the Lagrange multipliers satisfy the relation
$\lambda_i=\mu_i(c-v)$ (see \cite{Koi}).

One can verify that the construction goes through the Chaplygin reduction.
This property can be used in producing non-trivial nonholonomic geodesic flows
on $SO(n)$ which, after the $SO(n-1)$-reduction, give rise to
integrable systems on the sphere $S^{n-1}$.

In the case of Hamiltonian systems, under a similar reduction, the
Kovalevskaya and Goryachev--Chaplygin integrable cases of rigid body dynamics
result in integrable geodesic flows on $S^2$ that possess
 additional polynomial integrals of degree 4 and 3 in momenta respectively
(see \cite{BKF}).

\paragraph{Veselova problem with potentials.} Now let us go back to the $n$-dimensional
Veselova problem and
suppose that the $n$-dimensional rigid body is placed in an axisymmetric potential
force field $v=v(e_1)$ (recall that $\{e_1,\dots,e_n\}$ are redundant coordinates on $SO(n)$).
Then the equations of motion have the form
\begin{align}
&\frac{d}{dt}\left({\mathcal I}\omega\right) =[{\mathcal I}\omega ,\omega]
+\frac{\partial v}{\partial e_1} \wedge e_1+
\sum_{2\le p<q\le n}\lambda_{pq} \, e_p\wedge e_q , \nonumber \\
&\dot{e}_{i}+\omega e_{i} =0 , \qquad i=1,\dots, n ,
\label{Veselova_eq2}
\end{align}
togeteher with the constraints (\ref{Veselova_constraints}).

The potential is $SO(n-1)$--invariant and induces a well defined reduced
potential $V(q)$ on the sphere $S^{n-1}$. Here $V(q)\equiv
v(e_1)\vert_{e_1=q}$. The perturbed reduced system with the inertia
tensor (\ref{inertia_tensor}) has the same Chaplygin reducing
multiplier as the nonperturbed one. Therefore, in the new time
$\tau$, the reduced system becomes a natural mechanical system on
the sphere with the kinetic energy (\ref{L^*}) and the potential
$V(q)$.

Let $A_1<\dots<A_n$. It is known, that the most general separable potentials
compatible with the metric (\ref{L^*}) in the variables
$\{\lambda_1,\dots,\lambda_{n-1}\}$ have the form
\begin{equation}
V =\sum\limits_{k=1}^{n-1}\frac{\Delta_k}{\prod_{s\neq k}
\left( \lambda _{k}-\lambda _{s}\right) } ,
\label{potential}
\end{equation}
where $\Delta_k$ are functions of the variable $\lambda_k$ only
(see \cite{KBM}). Note that this potentials are of the same form
as the potentials compatible with the standard metric in the same
coordinates (e.g., see \cite{Wo}). Then, if $\Delta_k$ is a Laurent polynomial in
the variable $\lambda_k$, then the potential
(\ref{potential}) is a Laurent polynomial in the coordinates
variables $q_1,\dots,q_n$ (see, e.g.,  \cite{KBM, DJ, Wo}). In particular,
the reduced Veselova problem with potential
$$
V(q)=\alpha_1(A^{-1}q,q)+\alpha_2((A^{-1}q,A^{-1}q)-(A^{-1}q,q)^2)
+\sum_{i=1}^n\frac{\alpha_{i+2} }{q_i^2} ,
$$
$\alpha_i$ being arbitrary constants, is completely integrable.

Now assume that
$v(e_1)=\alpha_1(A^{-1}e_1,e_1)+\alpha_2((A^{-1}e_1,A^{-1}e_1)-(A^{-1}e_1,e_1)^2)$
and that the total energy is bigger than $\max_{SO(n)} v$. Let, as
above, $ds^2_\mathcal I$ be the left--invariant metric given by
the inertia operator (\ref{inertia_tensor}) and introduce the
Jacobi metric $ds^2_J=(c-v(e_1))ds^2_\mathcal I$. From the above
considerations and the Maupertuis principle we get the following
result.

\begin{thm}
The $SO(n-1)$-reduction of the the nonholonomic geodesic flow
of the metric $ds^2_J$ with the constraints
(\ref{Veselova_constraints}) is completely integrable.
The phase space $T^*S^{n-1}$ is almost everywhere foliated
by invariant $(n-1)$--dimensional Lagrangian tori with
nonuniform quasi--periodic dynamics.
\end{thm}

\paragraph{The Lagrange case.}
In general, the operator (\ref{inertia_tensor}) is not a physical inertia operator
of a multidimensional rigid body. However, by taking $A_1=\dots=A_{n-1}$, $A_n>A_1/2$
we get
\begin{equation*}
\mathcal I\omega=I\omega+\omega I, \quad I=\diag(I_1,\dots,I_1,I_n),
\quad I_1=\frac{A_1^2}{2\det A}, \quad I_n=\frac{A_1A_n}{\det A}-\frac{A_1^2}{2\det A} .
\end{equation*}
In this case the system (\ref{Veselova_eq2}) represents the  motion
of a symmetric rigid body under the nonholonomic constraints.

In the presence of the homogeneous gravitational force field in
the direction $e_1$ we have $v=Mg(C,e_1)$, where $g$ is the
gravitational constant, $M$ is the mass and $C=(C_1,\dots,C_n)$ is
the position of the center of mass of the body.
If the mass center is placed on the axis of the dynamical
symmetry, then $v=MgC_n e_{1n}$ and the system (\ref{Veselova_eq2}) represents
a multidimensional version of the Lagrange top (see \cite{Be}).

In the new time $\tau$, the reduced system is
completely integrable according to a non-commutative version of the Liouvilee theorem.
Appart from the Hamiltonian function, there are integrals arrising from the
$SO(n-1)$--symmetry of the system,
$$
q_i\tilde p_j-q_j \tilde p_i, \qquad 1\le i<j \le n-1.
$$
As a result, the reduced phase space $T^*S^{n-1}$ is foliated
by two-dimensional invariant tori.

Note that there is an another generalization of a heavy rigid body
(\cite{Ra2}), which is based on the generalization of the three--dimensional
Euler--Poisson equations to the Euler--Poisson equations on the
semi-direct product $so(n)\times so(n)$.

\section{L+R Systems}

\subsection{Definition and Invariant Measure of L+R Systems}

It appears that LR systems on a unimodular Lie group $G$ can be viewed as a
limit case of certain artificial systems on the same group, which also
possess an invariant measure. The latter systems do not have a
straightforward mechanical or geometric interpretation and arise
as a ``distortion'' of a geodesic flow on $G$ whose kinetic energy
is given by a sum of a left- and right-invariant metrics.

\paragraph{Geodesic flow on $G$ with L+R metric.}
In addition to the nondegenerate linear operator ${\cal I}$ defining the
left-invariant metric $(\cdot,\cdot)_{{\cal I}}$,
introduce a constant linear operator
$\Gamma^{0}:\; \g\rightarrow \g$ defining a
right-invariant metric $(\cdot,\cdot)_{\Gamma }$ on the
$n$-dimensional compact Lie group $G$: for any vectors $u, v\in T_{g} G$
we put $(u,v)_{\Gamma }=\langle u g^{-1},\Gamma ^{0} v g^{-1}\rangle$.
We take the sum of both metrics and consider the corresponding geodesic flow on
$G$ described by the  Lagrangian
$$
l(\omega, g) =\frac12 \langle \omega,{\cal I}\omega \rangle
+\frac12 \langle g \omega g^{-1},\Gamma^0 \, g\omega g^{-1} \rangle
\equiv \frac12 \langle \omega,{\cal I}\omega \rangle
+ \langle\omega,\Gamma (g) \omega\rangle,
$$
where $\Gamma (g)={Ad}_{g^{-1}}\Gamma^0 {Ad}_g$ and $Ad_g$ is regarded as a matrix
operator acting on $\mathfrak g$.

Suppose that the total inertia operator ${\cal B}(g)= {\cal I}+\Gamma (g)$ is
nondegenerate and positive definite on the whole group $G$. The geodesic motion on the group is described by the Euler--Poincar\'e  equations
\begin{equation}
\dot x= [ x, \omega] + g^{-1} \frac{\partial l}{\partial g},\qquad
x =\frac{\partial l}{\partial\omega}={\cal B}\omega,
\label{i5.3}
\end{equation}
together with the kinematic equation $\dot g=g\cdot\omega$.

In order to find explicit expression for $g^{-1} ({\partial l}/{\partial g})$,
we first note that for any $Y\in {\mathfrak g}$,
$$
\langle Y,g^{-1} ({\partial l}/{\partial g})\rangle = v_Y (l),
$$
where $v_Y$ is the left-invariant vector field on $G$ generated by $Y$.
Since the metric $(\cdot,\cdot)_{{\cal I}}$ is left-invariant, we
have
\begin{gather*}
v_Y (l) = \frac 12 v_Y (\langle\omega,\Gamma\omega\rangle)=
\frac12 \langle \omega,\Gamma{\rm ad\,}_{Y}\omega
+{\rm ad\,}_{Y}^T\Gamma\omega\rangle=
\langle\Gamma\omega,[Y,\omega]\rangle
= \langle Y, {\rm ad\,}_\omega\, \Gamma\, \omega \rangle.
\end{gather*}
As a result, $g^{-1} ({\partial l}/{\partial g})={\rm ad\,}_\omega\, \Gamma\, \omega$.

Also, in view of the definition of $\Gamma$, its evolution is given by $n\times n$
matrix equation
\begin{equation}
\dot\Gamma=\Gamma{\rm ad\,}_\omega+{\rm ad\,}_\omega^T\Gamma.
\label{i5.4}
\end{equation}
Note that for compact group we have ~${\rm ad\,}_\omega^T=-{\rm
ad\,}_\omega$, and~$\dot\Gamma=[\Gamma,{\rm ad\,}_\omega]$.

Equations (\ref{i5.3}), (\ref{i5.4}) form a closed system on the space
$\mathfrak g\times {\bf Symm}(n)$ with the coordinates $\omega_i, \Gamma_{ij}$,
$i\le j=1,\ldots,n$. Indeed, since $\cal B$ is nondegenerate,
the derivative $\dot\omega$ is uniquely defined from (\ref{i5.3}).

\paragraph{L+R systems.}
Now we modify equations (\ref{i5.3}) by rejecting the term
$g^{-1} ({\partial l}/{\partial g})$. As a result, we obtain
another system on the space $\mathfrak g\times {\bf Symm}(n)$
\begin{equation}
\frac d{dt}({\cal B}\omega)={\rm ad\,}_\omega^T \, {\cal B}\omega,\quad
\frac d{dt} \Gamma=\Gamma\, {\rm ad\,}_\omega+{\rm ad\,}_\omega^T \, \Gamma,
\qquad{\cal B}={\cal I}+\Gamma .   \label{i5.8}
\end{equation}
This is generally non a Lagrangian system, and, in contrast to equations
(\ref{i5.3}), (\ref{i5.4}), it possesses the ``momentum'' integral
~$\langle{\cal B}\omega,{\cal B}\omega\rangle$.
In view of the structure of the kinetic energy,
we shall refer to the system (\ref{i5.8}) as {\it L+R system} on $G$.

\begin{thm}\label{5.3} The L+R system (\ref{i5.8}) possesses
the kinetic energy integral $\frac 12\langle\omega,{\cal
B}\omega\rangle$ and an invariant measure ~$\mu\,d\omega_1
\wedge\cdots\wedge \omega_n\land
d\Gamma_{11}\wedge\cdots\wedge\Gamma_{nn}$
 with density
\begin{equation} \mu=\sqrt{\det ({\cal I}+\Gamma )}\, . \label{i5.9} \end{equation}
\end{thm}

\begin{remark}{\rm
The L+R systems can be also naturally considered on non-compact
groups. Then Theorem \ref{5.3} holds for unimodular groups as
well. Recall that the group $G$ is unimodular if tr ad$_{\omega}=0$.}
\end{remark}

\noindent{\it Proof of Theorem} \ref{5.3}.
First, replace $\frac d{dt}({\cal B}\omega)$ with $\mathcal B\dot \omega+\dot\Gamma\omega$.
Then, using (\ref{i5.4}) and the identity
ad$_{\omega }\omega =0,$ we can represent equations (\ref{i5.8}) in the form
\begin{equation}
{\cal B}\, \dot\omega= {\rm ad\,}_\omega^T\mathcal I \omega,\quad
\dot\Gamma= \Gamma{\rm ad\,}_\omega+{\rm ad\,}_\omega^T\Gamma.
\label{i5.12}
\end{equation}
Using this form, we compute
\begin{align*}
\frac d{dt} \langle\omega,{\cal B}\omega\rangle &=2 \langle\omega,{\cal B}\dot\omega\rangle +
  \langle\omega,\dot {\cal B}\omega\rangle \\
& = 2 \langle\omega,{\rm ad\,}_\omega^T\mathcal I \omega \rangle
+ \langle\omega,\Gamma{\rm ad\,}_\omega \omega +{\rm ad\,}_\omega^T\Gamma\omega \rangle =0.
\end{align*}
i.e., $\langle\omega,{\cal B}\omega\rangle$ is a first integral.

Next, divergence $\Delta $ of the phase flow of the system is calculated by the formula
\begin{equation}
\Delta=\sum_{i\le j}^n
\frac{\partial\dot{\Gamma}_{ij}}{\partial\Gamma_{ij}}+
\sum_{i=1}^n\frac{\partial\dot{\omega}_i}{\partial\omega_i}.
\label{i5.6}
\end{equation}
In view of (\ref{i5.4}),
the first sum equals
$
\sum_{i\le j}^n\left[({\rm ad\,}_\omega)_{jj}+({\rm ad\,}_\omega)_{ii} \right]=0.
$
Then we can write
$$
\Delta={\rm tr\,}({\cal B}^{-1}U),\qquad
U_{ij}=\frac{\partial(\ad^T_\omega \mathcal I\omega)_i}{\partial \omega_j}, \quad  i, j=1,\dots,n.
$$
As follows from the first equation in (\ref{i5.8}), here we can put
$U=\mathrm ad_{\mathcal I\omega}+\mathrm ad^T_\omega\mathcal I$.

In view of symmetry of ${\cal B}^{-1}$, the skew symmetric part of $U$ does not
contribute to the expression for $\Delta$. The symmetric part of $U$ has the
form
$$
U^+\equiv \frac12(U+U^T)=\frac12\left ({\rm ad\,}_\omega^T
({\cal B}-\Gamma)+({\cal B}-\Gamma)\, {\rm ad\,}_\omega\right).
$$
As a result, taking into account (\ref{i5.4}), we obtain
\begin{align*}
\Delta={\rm tr\,}({\cal B}^{-1}U^+) & =
\frac12{\rm tr\,}\left ({\cal B}^{-1}{\rm ad\,}_\omega^T{\cal B}+
{\rm ad\,}_\omega-{\cal B}^{-1}\dot\Gamma\right) \\
& =-\frac12{\rm tr\,}({\cal B}^{-1}\dot\Gamma)
= -\frac12{\rm tr\,}({\cal B}^{-1}\dot{\cal B}). \label{i5.14}
\end{align*}
Now, using the unimodularity condition $\mathrm{tr} \, \mathrm{ad}_{\omega}=0$ and
the well-known identity
\begin{equation}
\label{5.To}
\frac d{dt}\, \det{\cal B}=\det{\cal B}\;{\rm tr\,}({\cal B}^{-1}\dot{\cal B}),
\end{equation}
we conclude that $\mu =\sqrt{\det  {\cal B}}$ satisfies the Liouville equation
$\frac d{dt}(\ln  \mu )+\Delta =0,$ which establishes the theorem.

\paragraph{Chaplygin's sphere.} One of the best known examples of nonholonomic systems
with an invariant measure is the celebrated Chaplygin sphere. It described
a dynamically non-symmetric ball rolling without sliding on a horizontal plane. The center
of the mass is assumed to be at the geometric center. Under these condition the motion
is integrable (\cite{Ch1, Ch}).

It appears that a reduction of Chaplygin's sphere can be regarded as a $L+R$ system.
Namely, the original configuration space is $\R^2\times SO(3)$ and
the nonholonomic constraints define a $SE(2)$-invariant
three-dimensional distribution. Then one can regard the system as
an a Chaplygin system on the trivial bundle $\R^2\times SO(3)\to SO(3)$.
After the $\R^2$-reduction we obtain a system on $TSO(3)$, which,
written in the body frame, takes the following vector form \begin{gather}
\dot K=K\times\Omega, \qquad
K=J\Omega+ ma^2\Omega-ma^2 (\Omega,\gamma)\gamma \label{Chap}\\
\dot\alpha=\alpha\times\Omega,\quad
\dot\beta=\beta\times\Omega,\quad
\dot\gamma=\gamma\times\Omega, \nonumber
\end{gather}
where $J$, $a$, $m$, are the inertia operator, radius, and
mass of the ball respectively. Next, $\Omega$ is vector of the angular velocity and
$K$ is vector of the angular momentum at the contact point; $\alpha,\beta,\gamma$ are
unit vectors forming a fixed orthonormal frame in space,
$\gamma$ is assumed to be vertical vector.
The components of these vectors can be regarded as redundant coordinates on $SO(3)$.

Equations (\ref{Chap}) can be resolved with respect to $\dot \Omega$ to give
\begin{equation}
\label{chap_vector}
\begin{gathered}
{\cal I}\dot \Omega = {\cal I}\Omega \times \Omega
+\frac {ma^2}{ 1-ma^2(\gamma, {\mathcal I}^{-1}\gamma) }
({\cal I}\Omega \times \Omega , {\mathcal I}^{-1}\gamma)\, \gamma, \\
\dot\alpha=\alpha\times\Omega, \quad \dot\beta=\beta\times\Omega, \quad
\dot\gamma=\gamma\times\Omega .
\end{gathered}
\end{equation}

After usual identification of the Lie algebras $(\R^3,\times)$ and $(so(3),[\cdot,\cdot])$,
the system (\ref{Chap}) can be seen
as a L+R system (\ref{i5.8}) on $SO(3)$ with left invariant metric given by the
${\mathcal I} : so(3)\to so(3)$ and right invariant
degenerate operator $\Gamma=-ma^2 \gamma\otimes \gamma$. According to
Theorem \ref{5.3}, in the space $(\Omega, \gamma)$ the above equations have an
invariant measure with density $\sqrt{\det (\mathcal I-ma^2 \gamma\otimes \gamma)}$.
Up to a constant factor, it equals $\sqrt{ 1-ma^2(\gamma, {\mathcal I}^{-1}\gamma )}$,
the expression given by Chaplygin in \cite{Ch1}.

Note that, in contrast to what was belived earlier,
 Chaplygin's sphere cannot be represented as an LR system on the group $SE(3)$
(see \cite{Schneider}).

It is interesting that equations (\ref{Chap}) are Hamiltonian
with respect to a certain nonlinear brackets (see Borisov and
Mamaev \cite{BM, BM2}).

\subsection{The spherical Support} 
\begin{figure}[h,t]
\begin{center}
\includegraphics[width=.7\textwidth]{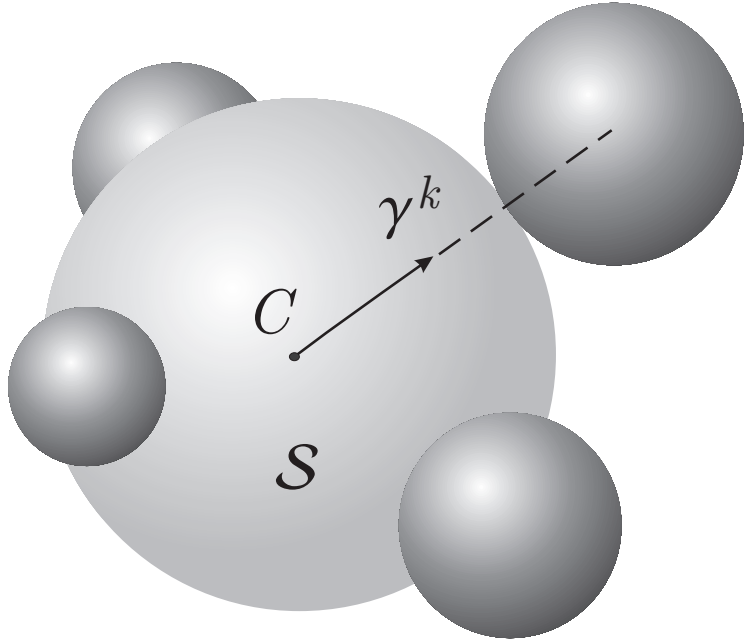}
\caption{\footnotesize The spherical support}
\label{st.fig}
\end{center}
\end{figure}

The Chaplygin sphere admits an integrable
generalization on the configuration space $SO(3)$.
Namely, consider the motion of a dynamically nonsymmetric ball $\cal S$ with
the unit radius around its fixed center. Suppose that the ball
touches $N$ arbitrary dynamically symmetric balls whose centers
are also fixed, and there is no sliding at the contacts points. We
call this mechanical construction {\it the spherical support}
(\cite{Fed_nonholonomic, Fe1}, see Figure \ref{st.fig}).

Let $\Omega\in {\mathbb R}^3$ and $J\, :\,{\mathbb R}^3 \to
{\mathbb R}^3$ be respectively the angular velocity vector and the
inertia tensor of the ball $\cal S$ in a frame attached to the ball.
Next, let ${\bf w} ^{k}\in {\mathbb R}^{3}, D_{k},\rho_{k}\in
{\mathbb R}$ be the angular velocity, the central inertia moment
and the radius of the $k$th peripheral ball, $\gamma^{k}$ be the
fixed {\it unit\/} vector directed from the center $C$ of the ball
$\cal S$ to the point of contact with the $k$th ball, $R^{k}$ be the
reaction force at this point acting on $\cal S$. Then the equations of
motion of the total mechanical system can be written in the form
\begin{equation}
J\dot\Omega+ \Omega\times J\Omega =\sum_{k=1}^N \gamma^k \times R^k,\quad
D_k\dot{\bf w}^k= -\rho_k \gamma^k \times R^k,\qquad  k=1,\dots,N,
\label{i5.15}
\end{equation}
where, as above, $\times$ denotes the standard vector product in
${\mathbb R}^3$.
Note that the first equation is taken in the moving frame,
whereas the other equations are taken in {\it a fixed\/} frame.

The reaction forces are due to nonholonomic constraints expressing
the absence of sliding at the contact points. This means that
velocity of the point of contact of the ball $\cal S$ with the $k$th ball,
$\Omega\times\gamma^k$, is the same as the
velocity of the corresponding point on the $k$th ball, i.e.,
${\bf w}^k\times(-\rho_k\gamma^k)$.  Multiplying the velocities
by the vectors $\gamma^k$ we obtain the constraints in form
$$
\rho_k ({\bf w}^k,\gamma^k)\gamma^k -\rho_k {\bf w}^k =\Omega
-(\Omega,\gamma^k)\gamma^k, \quad k=1,\dots,N.
$$
By differentiating the constraints {\it in the fixed frame\/}
and taking into account $\dot\gamma^{k}=0$, $({\bf w}^k,\gamma^k)={\rm const}$,
we get
 $$
\rho_k \dot{\bf w}^k = -\dot\Omega +(\dot\Omega,\gamma^k)\gamma^k
$$
and, in view of (\ref{i5.15}),
$$
\gamma^k \times R^k=-\frac{D_k}{\rho_k^2}[(\dot\Omega,\gamma^k)\gamma^k -\dot\Omega].
$$
Substituting this into the first equation in (\ref{i5.15}) and
using the fact that the time derivatives of $\Omega$ in the moving
and the fixed frames are the same, we obtain
\begin{gather}
{\cal I}\dot\Omega+ \Omega\times J\Omega =-\Gamma \dot\Omega,\label{i5.16} \\
{\cal I}= J -\sum_{k=1}^N  \frac{D_k}{\rho_k^2} {\bf I}, \quad
\Gamma=\sum_{k=1}^N  \frac{D_k}{\rho_k^2} \gamma^k\otimes\gamma^k ,
\label{i5.17}
\end{gather}
where ${\bf I}$ is the $3\times 3$ identity matrix and
$\Gamma$ is the $3\times 3$ symmetric matrix, which is fixed in
the space. For $N\ge 3$ and a general location of the peripheral
balls, it is nondegenerate, hence its components can be regarded
as redundant coordinates on the group $SO(3)$. Since the evolution
of $\gamma^k$ in the moving frame is described by the Poisson
equations $\dot\gamma^k =\gamma^k \times\Omega$, from
(\ref{i5.17}) we have
\begin{equation}
\dot \Gamma =[\Gamma, \omega],  \label{PG}
\end{equation}
where $\omega\in so(3)$ is the $3\times 3$ skew-symmetric matrix
such that $\omega_{ij}=\varepsilon_{ijk} \Omega_k$.

Now we consider the motion of the central
ball $\cal S$ only. As follows from (\ref{i5.16})-(\ref{PG}), equations of motion can
be represented in the form of an L+R system on the group $SO(3)$,
\begin{gather} \label{i5.18}
\dot K =K\times \Omega,\quad \dot \Gamma =[\Gamma, \omega], \\
K=({\cal I}+\Gamma)\Omega \in {\mathbb R}^3 . \nonumber
\end{gather}
Notice that from here $\dot\Omega$ can be uniquely expressed in terms
of the components of $\Omega, \Gamma$, hence (\ref{i5.18}) represents a closed
system of differential equations.
One can say that it describes the free rotation of a ``generalized
Euler top'', whose tensor of inertia is a sum of two components: one is fixed in
the body and the other one is fixed in the space.

\begin{thm} The spherical support system \textup{(\ref{i5.18})} is integrable by
the Euler--Jacobi theorem, and its generic invariant manifolds are two-dimensional tori.
\end{thm}

Indeed, we can put $\Gamma =a\alpha \otimes \alpha +b\beta \otimes
\beta +c\gamma \otimes \gamma$, where $\alpha,\beta,\gamma$ are
unit vectors forming a fixed orthonormal frame in the space and
$a,b,c$ are some constants, which can be uniquely determined from
(\ref{i5.17}). Then the matrix equation in (\ref{i5.18}) can be
replaced by the vector equations
\begin{equation} \label{PP}
\dot\alpha=\alpha\times\Omega,\quad
\dot\beta=\beta\times\Omega,\quad
\dot\gamma=\gamma\times\Omega .
\end{equation}

From the form of equations (\ref{i5.18}), (\ref{PP}) we
immediately obtain four first integrals
\begin{gather*}
(K,K),\quad (K,\alpha)=({\cal I}_a\omega,\alpha),\quad (K,\beta)=
({\cal I}_b\omega,\beta),\quad (K,\gamma)=({\cal I}_c\omega,\gamma), \\
{\cal I}_a={\cal I}+a{\bf I},\quad {\cal I}_b={\cal I}+b{\bf I}, \quad {\cal I}_c={\cal I}+c{\bf I},
\end{gather*}
of which any three integrals are independent. In addition, the system has
trivial geometric integrals
\begin{gather*}
(\alpha,\alpha)=1,\quad (\beta,\beta)=1,\quad (\gamma,\gamma)=1,\\
(\alpha,\beta)=0,\quad (\alpha,\gamma)=0,\quad (\beta,\gamma)=0
\end{gather*}
and the kinetic energy integral
$$\frac 12(\omega, ({\cal I}+\Gamma)\omega)=\frac12(\omega,{\cal I}\omega)+
\frac a2(\omega,\alpha)^2+\frac b2(\omega,\beta)^2+\frac c2(\omega,\gamma)^2.$$
Next, according to Theorem \ref{5.3}, the system also possesses an invariant
measure with density
\begin{gather}
\mu=\sqrt{\det ({\cal I}+\Gamma)}=
\sqrt{\det{\cal I}}\bigg (1+a (\alpha,{\cal I}^{-1}\alpha)+b (\beta,{\cal I}^{-1}\beta)
+c (\gamma,{\cal I}^{-1}\gamma)+ \nonumber \\
+\frac{bc}{\det{\cal I}}(\alpha,{\cal I}\alpha)+
\frac{ac}{\det{\cal I}}(\beta,{\cal I}\beta)+
\frac{ab}{\det{\cal I}}(\gamma,{\cal I}\gamma)+\frac{abc}{\det{\cal I}}
\bigg)^{1/2}.
\label{measure2}
\end{gather}
This together implies the integrability by the Euler--Jacobi theorem. Notice
that for the case of only one peripheral ball, the
L+R system (\ref{i5.18}) has the same form as Chaplygin's ball system (\ref{Chap}).

\subsection{Limits of L+R Systems}
As mentioned above, a nonholonomic LR system on a Lie group $G$
can be obtained as a limit case of a certain L+R system on this
group. Indeed, suppose that the operator $\Gamma\; :\; \mathfrak
g\to \mathfrak g$ defining a right-invariant metric on $G$ is {\it
degenerate} and has the form
\begin{equation}
\Gamma=\epsilon({\alpha}^1\otimes{\alpha}^1+\cdots+{\alpha}^{\rho}
\otimes{\alpha}^{\rho}),
\qquad \rho <n,\quad D={\rm const}>0,
\label{i5.26}
\end{equation}
where, as in (\ref{Rconstr}),
${\alpha}^1, \dots, {\alpha}^{\rho}$
are orthonormal right-invariant vector fields
$\alpha^i=g^{-1}\cdot a^i \cdot g$, $a^i=\mbox{const}\in g$, generating a right-invariant distribution  $D$ on $TG$.

Now consider the L+R system (\ref{i5.8}) on the space
$(\omega,{\alpha}^1,\dots,{\alpha}^{\rho})$. In view of (\ref{i5.12}),
it can be represented in form
\begin{equation} \label{resolved}
{\cal I} \dot\omega= {\cal I} ({\cal I}+\Gamma)^{-1} {\rm ad\,}_\omega^T\mathcal I \omega,\quad
\dot\Gamma=\Gamma{\rm ad\,}_\omega+{\rm ad\,}_\omega^T\Gamma.
\end{equation}
 Then the following theorem holds (see \cite{Fe1}).

\begin{thm}
\begin{description}
\item{1).} As $\epsilon\rightarrow \infty $, equations (\ref{resolved})
transform to the Euler--Lagrange equations with multipliers (\ref{lr3}) and constraints
(\ref{Rconstr}), where $x={\cal I}\omega$.
\item{2).} The density $\sqrt{\det {\cal B}}/\sqrt{\epsilon}$ of the invariant measure
of the L+R
system tends to the density (\ref{density}) of the LR system multiplied by a
constant factor.
\end{description}
\end{thm}

Note that as $\epsilon\rightarrow \infty $, the original equations (\ref{i5.8})
become singular. For this reason, before taking the limit they must be transformed
to the form (\ref{resolved}).

As an illustration, consider the following $L+R$ system on $SO(3)$
(we use the usual vector notation):
\begin{equation}
\mathcal I\dot\Omega=\mathcal I(\mathcal I+\epsilon\gamma\otimes\gamma)^{-1}
(\mathcal I\Omega\times\Omega),  \qquad \dot\gamma=\gamma\times\Omega,
\label{L+R}
\end{equation}
which formally coisides with the Chaplygin sphere system (\ref{chap_vector}) if we set $\epsilon=ma^2$.
It can be easily verified that
$$
\lim_{\epsilon \to \infty}
(\mathbf{I}+\epsilon \gamma \otimes \mathcal I^{-1}\gamma )^{-1}=
\mathbf{I}-\frac{1}{(\mathcal I^{-1}\gamma,\gamma)}\gamma\otimes\mathcal I^{-1}\gamma.
$$
Therefore, as $\epsilon$ tends to infinity, the system (\ref{L+R})
transforms to the Veselova rigid body problem (\ref{3.1}).

\subsection*{Acknowledgments}

We would like to thank Alexey Bolsinov, Anatoly Fomenko and Andrey Oshemkov
for the kind invitation for writing this paper.

The first author (Yu.F.) acknowledges the support of grant BFM 2003-09504-C02-02 of
Spanish Ministry of Science and Technology.

The second author (B.J.) was partially supported by the Serbian Ministry of
Science and Technology, Project 1643 (Geometry and Topology of Manifolds and
Integrable Dynamical Systems).

\end{document}